# THIN WAVEGUIDES WITH ROBIN BOUNDARY CONDITIONS

GUY BOUCHITTÉ, LUÍSA MASCARENHAS, AND LUÍS TRABUCHO

ABSTRACT. We consider the Laplace operator in a thin three dimensional tube with a Robin type condition on its boundary and study, asymptotically, the spectrum of such operator as the diameter of the tube's cross section becomes infinitesimal. In contrast with the Dirichlet condition case [2], we evidence different behaviors depending on a symmetry criterium for the fundamental mode in the cross section. If that symmetry condition fails, then we prove the localization of lower energy levels in the vicinity of the minimum point of a suitable function on the tube's axis depending on the curvature and the rotation angle. In the symmetric case, the behavior of lower energy modes is shown to be ruled by a one dimensional Sturm-Liouville problem involving an effective potential given in explicit form.

## 1. Introduction

In a previous paper [2], the authors presented a new variational approach by Γ-convergence in order to study the asymptotic behavior of the spectral problem for the Laplace operator with homogeneous Dirichlet boundary conditions in a tube of infinitesimal thickness. The limit problem arising from a 3D-1D reduction analysis was shown to be characterized by a 1D-effective potential depending explicitly on the local curvature and torsion. From there, very interesting effects on the energy levels could be evidenced in terms of the geometrical characteristics of the thin domain, in a way which was complementary to many results in the literature, as for instance in [3], [6], [7].
In the present paper we perform the same analysis for the case of the Laplace operator with Robin boundary conditions, more precisely, we consider the eigen problem:

$$\begin{cases} -\Delta u_\varepsilon = \lambda^\varepsilon u_\varepsilon, & \text{in } \Omega_\varepsilon, \\ \frac{\partial u_\varepsilon}{\partial n_\varepsilon} + \gamma_\varepsilon u_\varepsilon = 0, & \text{on } \partial\Omega_\varepsilon. \end{cases} \quad (1.1)$$

where $\varepsilon$ is a small positive parameter, $\Omega_\varepsilon \subset \mathbb{R}^3$ is a thin and long domain generated by a cross section $\omega_\varepsilon = \varepsilon\,\omega$ (being $\omega$ a fixed subset of $\mathbb{R}^2$) which rotates along a curve through an angle $\alpha(s)$ with respect to the Frenet frame. Here the function $\gamma_\varepsilon$ is a suitable scaled real coefficient in $L^\infty(\partial\Omega_\varepsilon, \mathbb{R}^+)$. In terms of local coordinates $x = \Psi_\varepsilon(s, y)$ with $(s, y) \in [0, L] \times \partial\omega$, (see (2.6) in Section 2), it has the form

$$\gamma_\varepsilon(x) = \begin{cases} \frac{\gamma(y)}{\varepsilon}, & \text{for } (y,s) \in ]0,L[\times \partial\omega, \\ \gamma_0(y), & \text{for } (y,s) \in \{L\} \times \omega, \\ \gamma_L(y), & \text{for } (y,s) \in \{L\} \times \omega, \end{cases} \quad (1.2)$$

where function $\gamma \in L^\infty(\partial\omega; \mathbb{R}^+)$ is a weight for the Robin condition on the lateral part of the thin tube $\Omega_\varepsilon$ whereas $\gamma_0, \gamma_L \in L^\infty(\omega; \mathbb{R}^+)$ are associated with the Robin condition we set on the two bases. Notice that the Dirichlet case studied in [2] can be formally recovered by taking $\gamma, \gamma_0, \gamma_L = +\infty$. However the situation is quite different here and the asymptotic analysis as $\varepsilon \to 0$ of the eigenvalue problem (1.1) under the scaling given in (1.2) reveals an important novelty. Indeed, two rather distinct situations will occur depending on the geometric constant vector $\rho_0 := \frac{1}{2} \int_{\partial \omega} u_0^2 \, n \, d\sigma$ where $u_0$ is the fundamental mode in the cross section $\omega$ with exterior unitary normal $n$. If $\rho_0$ vanishes, which is the case when subset $\omega$ and function $\gamma$ present enough symmetry, then the lower level eigenmodes are propagating along the central curve and are characterized through a suitable 1D spectral problem with a potential weighted by local torsion and curvature; thus the situation is similar to the Dirichlet case treated in [2].

In contrast, if $\rho_0$ is a non zero vector, then a localization phenomenon takes place in the vicinity of the minimum point of a suitable function on the central curve depending on the curvature and on the rotation angle. In that case we show that the low level eigenmodes behave, after blow-up, like the eigenfunctions of a 1D-harmonic oscillator. Let us notice that similar effects have been pointed out recently in [1] where narrow strips in $\mathbb{R}^2$ are considered whose thickness presents a strict global maximizer. Two dimensional waveguides with mixed Dirichlet and Neumann conditions have been also considered in [8],[9].

In Section 2, after introducing the geometry of the waveguide and the scaling, we present our asymptotic variational approach and some preliminary results. In particular, we give a perturbation result for the fundamental eigenvalue in the cross section. In Section 3 we study the *symmetric* case ($\rho_0 = 0$) and prove the spectral convergence to a 1D limit Sturm-Liouville problem. The non symmetric case $\rho_0 \neq 0$ is considered in Section 4. We prove the localization of the lower energy levels and evidence a gap between them, blowing up like $\varepsilon^{-1/2}$ as $\varepsilon \to 0$.

## 2. Definitions and preliminary results

**2.1. Geometry of the domain.** Let $r : s \in [0, L] \to r(s) \in \mathbb{R}^3$ be a simple $C^2$ curve in $\mathbb{R}^3$ parametrized by the arc length parameter $s$. Denoting by $T$ its tangent vector and assuming that $T'(s) \neq 0$ for every $s \in [0, L]$, we may define the usual Frenet system $(T, N, B)$ through the following expressions:

$$T = \frac{dr}{ds} = r' \quad (\|r'\|_{\mathbb{R}^3} = 1); \quad N = T'/\|T'\|_{\mathbb{R}^3}; \quad B = T \times N.$$

Denote by $k : s \in [0, L] \to k(s) \in \mathbb{R}$ and by $\tau : s \in [0, L] \to \tau(s) \in \mathbb{R}$, the curvature and torsion functions associated with the curve, respectively. They are functions in $L^\infty(0, L)$ and they satisfy the Frenet formulas:

$$T' = k \, N; \quad N' = -k \, T + \tau \, B; \quad B' = -\tau \, N. \tag{2.1}$$

It is clear from (2.1) that the plane defined by $(N(s), B(s))$ rotates around $T(s)$, as $s$ moves along $[0, L]$. On the contrary, if we consider the Tang system $(T(s), X(s), Y(s))$ for $X$ and $Y$ satisfying

$$X' = \lambda \, T; \quad Y' = \mu T; \quad T' = -\lambda \, X - \mu \, Y, \tag{2.2}$$

where $\lambda$ and $\mu$ are functions of the arclength parameter $s$, the plane defined by $(X(s), Y(s))$ does not rotate around $T(s)$. It is easy to check from (2.1) and (2.2) that the velocity of the rotation $\alpha_0(s)$ of $(N(s), B(s))$ with respect to $(X(s), Y(s))$ at each point $s \in [0, L]$ satisfies $\alpha_0'(s) = \frac{d\alpha_0}{ds} = -\tau(s)$; we also obtain that $\lambda = -k \cos \alpha_0$, $\mu = k \sin \alpha_0$ (see [2]).
The twisted thin domain on which we will study the energy levels of problem (1.1) will be described by a rotation function $\alpha \in L^\infty(0, L)$. Let us define

$$\begin{aligned} N_\alpha(s) &:= \cos \alpha(s) \, N(s) + \sin \alpha(s) \, B(s) = \cos(\alpha - \alpha_0)(s) \, X(s) + \sin(\alpha - \alpha_0)(s) \, Y(s), \\ B_\alpha(s) &:= -\sin \alpha(s) \, N(s) + \cos \alpha(s) \, B(s) = -\sin(\alpha - \alpha_0)(s) \, X(s) + \cos(\alpha - \alpha_0)(s) \, Y(s). \end{aligned} \tag{2.3}$$

Then, given $\omega \subset \mathbb{R}^2$ an open bounded simply connected subset of $\mathbb{R}^2$, we define for every small parameter $\varepsilon > 0$

$$\Omega_\varepsilon := \left\{ x \in \mathbb{R}^3 : x = r(s) + \varepsilon\, y_1\, N_\alpha + \varepsilon\, y_2\, B_\alpha,\ s \in [0,L],\ y = (y_1, y_2) \in \omega \right\}. \qquad (2.4)$$

The diameter of the cross section of the domain $\Omega_\varepsilon$ is of infinitesimal order $\varepsilon$ (in particular much smaller than the length $L$). Moreover, the local torsion at every point of the central curve $r(s)$ is measured by the parameter $\tilde{\tau} := \tau + \alpha'$, i.e. by the velocity of rotation of the cross section with respect to the Tang system.

## 2.2. Variational formulation on a fixed domain.

We start from the variational formulation of problem (1.1):

$$\int_{\Omega_\varepsilon} \nabla u_\varepsilon\, \nabla w + \int_{\partial\Omega_\varepsilon} \gamma_\varepsilon\, u_\varepsilon\, w = \lambda^\varepsilon \int_{\Omega_\varepsilon} u_\varepsilon\, w \qquad \text{for all } w \in H^1(\Omega_\varepsilon),$$

to which we associate the quadratic energy functional defined in $H^1(\Omega_\varepsilon)$ by:

$$F_\varepsilon(w) := \int_{\Omega_\varepsilon} |\nabla w|^2\, dx + \int_{\partial\Omega_\varepsilon} \gamma_\varepsilon\, |w|^2\, d\sigma. \qquad (2.5)$$

As usual in dimension reduction analysis, it is convenient to deal with an equivalent formulation on a fixed domain $Q_L :=\,]0, L[\times\omega$. In relation with (2.3) and (2.4), we consider for each $\varepsilon > 0$ the following transformation

$$\begin{aligned} \psi_\varepsilon\ &:\ \overline{Q}_L\ \longrightarrow\ \overline{\Omega}_\varepsilon, \\ (s, y)\ &=\ (s, (y_1, y_2)) \mapsto x = r(s) + \varepsilon\, y_1 N_\alpha + \varepsilon\, y_2 B_\alpha. \end{aligned} \qquad (2.6)$$

Accordingly, to every element $u \in H^1(\Omega_\varepsilon)$, we associate $v \in H^1(Q)$ defined by

$$v(s, (y_1, y_2)) := u(\psi_\varepsilon(s, (y_1, y_2))). \qquad (2.7)$$

We write the gradient of $v$ in the form $(v', \nabla_y v)$, being $v'$ the derivative with respect to $s \in [0, L]$. In order to compute the Dirichlet energy of $u$ on $\Omega_\varepsilon$, we introduce

$$\tilde{\tau} := \tau + \alpha',\quad \beta_\varepsilon(s, y) := 1 - \varepsilon k(s)(z_\alpha \cdot y),\quad z_\alpha := (\cos\alpha, -\sin\alpha),\quad z_\alpha^\perp := (\sin\alpha, \cos\alpha),\ (2.8)$$

where $z_\alpha \cdot y$ represents the inner product in $\mathbb{R}^2$ of $z_\alpha$ and $y$.
Then, after some computations, we get

$$\nabla\psi_\varepsilon = \begin{pmatrix} \beta_\varepsilon & 0 & 0 \\ -\varepsilon\tilde{\tau}(z_\alpha^\perp \cdot y) & \varepsilon\cos\alpha & -\varepsilon\sin\alpha \\ \varepsilon\tilde{\tau}(z_\alpha \cdot y) & \varepsilon\sin\alpha & \varepsilon\cos\alpha \end{pmatrix},\quad \det\nabla\psi_\varepsilon = \varepsilon^2 \beta_\varepsilon,$$

$$\nabla\psi_\varepsilon^{-1} = \begin{pmatrix} \frac{1}{\beta_\varepsilon} & 0 & 0 \\ \frac{\tilde{\tau} y_2}{\beta_\varepsilon} & \frac{\cos\alpha}{\varepsilon} & \frac{\sin\alpha}{\varepsilon} \\ \frac{-\tilde{\tau} y_1}{\beta_\varepsilon} & \frac{-\sin\alpha}{\varepsilon} & \frac{\cos\alpha}{\varepsilon} \end{pmatrix}.$$

Thus we have

$$\int_{\Omega_\varepsilon} (|\nabla u(x)|^2) \, dx = \int_0^L \int_\omega (|\nabla v(s,y) \, \nabla \psi^{-1}(s,y)|^2) \, \varepsilon^2 \, \beta_\varepsilon(s,y) \, dy \, ds$$
$$= \varepsilon^2 \int_0^L \int_\omega \left[ \frac{1}{\beta_\varepsilon} \left| v' + (\nabla_y v \cdot R \, y) \tilde{\tau} \right|^2 + \frac{\beta_\varepsilon}{\varepsilon^2} \left( |\nabla_y v|^2 \right) \right] dy \, ds, \tag{2.9}$$

where $R$ is the clockwise rotation matrix $\begin{pmatrix} 0 & 1 \\ -1 & 0 \end{pmatrix}$.

Let now $x \in \Gamma_\varepsilon = ]0,L[ \times \partial \omega_\varepsilon$ and, representing by $t$ the local tangential coordinate along the oriented boundary of $\omega$, define, for $y = y(t)$, $\dot{y} := \dfrac{dy}{dt}$. We have

$$\frac{\partial x}{\partial s} \times \frac{\partial x}{\partial t} = \begin{vmatrix} T & N & B \\ \beta_\varepsilon & -\varepsilon \tilde{\tau}(z_\alpha^\perp \cdot y) & \varepsilon \tilde{\tau}(z_\alpha \cdot y) \\ 0 & \varepsilon(\dot{y} \cdot z_\alpha) & \varepsilon(\dot{y} \cdot z_\alpha^\perp) \end{vmatrix}$$

and, consequently,

$$\frac{1}{\varepsilon} \left\| \frac{\partial x}{\partial s} \times \frac{\partial x}{\partial t} \right\| = \sqrt{\beta_\varepsilon^2 + \varepsilon^2 \tilde{\tau}^2 (y \cdot \dot{y})^2} = \beta_\varepsilon + \varepsilon^2 r_\varepsilon, \tag{2.10}$$

where, as can be checked by (2.8) and Taylor expansion of the square root, the function $r_\varepsilon(s,y)$ satisfies

$$r_\varepsilon \geq 0 \quad \text{and} \quad \left| r_\varepsilon - \frac{\tilde{\tau}^2}{2}(y \cdot \dot{y})^2 \right| \leq C \varepsilon. \tag{2.11}$$

From (2.10), (2.11) and the fact that $\gamma \in L^\infty(0,L)$, it follows that

$$\left| \int_{\Gamma_\varepsilon} \gamma_\varepsilon \, |u|^2 \, d\sigma_\varepsilon(x) - \int_0^L \int_{\partial \omega} \gamma \, |v|^2 \left( \beta_\varepsilon + \frac{\varepsilon^2}{2} \tilde{\tau}^2 (y \cdot \dot{y})^2 \right) d\sigma \, ds \right| \leq C \varepsilon^3 \int_0^L \int_{\partial \omega} |v|^2 \, d\sigma \, ds. \tag{2.12}$$

On the other hand, if $x \in \Sigma_\varepsilon := \{0, L\} \times \omega_\varepsilon$, then $\left\| \frac{\partial x}{\partial y_1} \times \frac{\partial x}{\partial y_2} \right\| = \varepsilon^2$ and we get

$$\int_{\Sigma_\varepsilon} \gamma_\varepsilon \, |u|^2 \, d\sigma_\varepsilon(x) = \varepsilon^2 \int_\omega \left( \gamma_0 \, |v(0,y)|^2 + \gamma_L \, |v(L,y)|^2 \right) dy. \tag{2.13}$$

Let us define the functional $\tilde{F}_\varepsilon : H^1(Q_L) \to \mathbb{R}$ by setting

$$\frac{1}{\varepsilon^2} \tilde{F}_\varepsilon(v) := \int_0^L \int_\omega \frac{1}{\beta_\varepsilon} \left| v' + (\nabla_y v \cdot R \, y) \tilde{\tau} \right|^2 dy \, ds$$
$$+ \int_0^L \frac{\tilde{\tau}^2}{2} \left( \int_{\partial \omega} \gamma \, |v|^2 (y \cdot \dot{y})^2 \, d\sigma \right) ds$$
$$+ \int_\omega \left( \gamma_0 \, |v(0,y)|^2 + \gamma_L \, |v(L,y)|^2 \right) dy$$
$$+ \frac{1}{\varepsilon^2} \int_0^L \left[ \int_\omega \beta_\varepsilon |\nabla_y v|^2 dy + \int_{\partial \omega} \beta_\varepsilon \, \gamma \, |v|^2 d\sigma \right] ds \tag{2.14}$$

Then, recalling (2.7) and collecting (2.9), (2.12) and (2.13), we obtain, for small $\varepsilon$, the following estimate:

$$|F_\varepsilon(u) - \tilde{F}_\varepsilon(v)| \leq C\,\varepsilon^3\,\|v\|^2_{H^1(Q_L)}. \tag{2.15}$$

## 2.3. Perturbed problem in the cross section.
In view of the last term appearing in (2.14), an important step is to understand the behavior as $\varepsilon \to 0$ of the following minimal Rayleigh quotient in each cross section $\{s\} \times \omega$:

$$m_\varepsilon(s) := \inf_{v \in H^1(\omega),\, v \neq 0} \frac{\int_\omega \beta_\varepsilon(s,y)\,|\nabla_y v|^2\,dy + \int_{\partial\omega} \beta_\varepsilon(s,y)\,\gamma\,|v|^2\,d\sigma}{\int_\omega \beta_\varepsilon(s,y)\,|v|^2\,dy}. \tag{2.16}$$

Recalling that $\beta_\varepsilon(s,y) = 1 - \varepsilon \xi(s) \cdot y$, to each $\xi \in \mathbb{R}^2$ we associate the following perturbed spectral problem in $H^1(\omega)$:

$$\begin{cases} -\text{div}\Big([1 - \xi \cdot y]\,\nabla u\Big) = \lambda\,[1 - \xi \cdot y]\,u, & \text{in } \omega, \\ \dfrac{\partial u}{\partial n} + \gamma u = 0, & \text{on } \partial\omega. \end{cases} \tag{2.17}$$

For small values of $|\xi|$, the related operator is positive self-adjoint with compact resolvent. We denote by $\Lambda_0(\xi)$ the fundamental eigenvalue of (2.17). It is given by the following minimum problem:

$$\Lambda_0(\xi) = \inf_{v \neq 0} \left\{ \frac{\int_\omega (1-\xi \cdot y)\,|\nabla v|^2 dy + \int_{\partial\omega}(1-\xi \cdot y)\,\gamma\,|v|^2\,d\sigma}{\int_\omega (1-\xi \cdot y)|v|^2\,dy}\,, v \in H^1(\omega) \right\}, \tag{2.18}$$

Then, we observe that

$$m_\varepsilon(s) = \Lambda_0\left(\varepsilon\,k(s)\,z_{\alpha(s)}\right). \tag{2.19}$$

Therefore, it is worth studying the behavior of function $\Lambda_0(\xi)$ in a neighbourhood of $\xi = 0$. Let $(\lambda_0, u_0)$ be the first eigenpair of the Robin-Laplace operator in $\omega$, i.e.,

$$\begin{cases} -\Delta u_0 = \lambda_0\,u_0, & \text{in } \omega, \\ \dfrac{\partial u_0}{\partial n} + \gamma u_0 = 0, & \text{on } \partial\omega, \\ u_0 > 0, \quad \int_\omega u_0^2 = 1. \end{cases} \tag{2.20}$$

We associate with $u_0$ two vectors (which depend only on $\omega$ and $\gamma$):

$$\rho_0 := \frac{1}{2}\int_{\partial\omega} u_0^2 n\,d\sigma \quad \text{and} \quad y_0 := \int_\omega u_0^2\,y\,dy. \tag{2.21}$$

We obviously have that $\Lambda_0(0) = \lambda_0$ which is strictly positive since we took for $\gamma(s)$ a non negative function. Moreover, by Krein-Rutman's Theorem, $\lambda_0$ is a simple eigenvalue for (2.20) and the associated eigenvector $u_0$ can be chosen to be positive on $\omega$.

We notice that $u_0$ is orthogonal in $L^2(\omega)$ to all components of the vector function $\nabla u_0 - \rho_0\,u_0$ being $\rho_0$ given by (2.21). Indeed, by integration by parts, we have:

$$2\int_\omega u_0(\nabla u_0 - \rho_0\,u_0)\,dy = \int_\omega \left(\nabla\left(u_0^2\right) - 2\rho_0 u_0^2\right)\,dy = \int_{\partial\omega} u_0^2 \cdot n\,d\sigma - 2\rho_0 = 0.$$

Thus, by Freedholm's alternative, for every $\xi = (\xi_1, \xi_2) \in \mathbb{R}^2$, there exists a unique solution $\chi_\xi$ of

$$\begin{cases} -\Delta\chi_\xi - \lambda_0\,\chi_\xi = -\xi\cdot\nabla u_0 + \xi\cdot\rho_0\,u_0 & \text{in } \omega, \\ \frac{\partial\chi_\xi}{\partial n} + \gamma\,\chi_\xi = 0, & \text{on } \partial\omega, \\ \int_\omega \chi_\xi\,u_0\,dy = 0. \end{cases} \quad (2.22)$$

By linearity, we have $\chi_\xi = \xi_1\chi_1 + \xi_2\chi_2$ where the shape functions $\chi_i$ are solutions for $\xi = e_i$, $i = 1, 2$. Setting $\chi := (\chi_1, \chi_2)$ and denoting by $I_2$ the $2 \times 2$ identity matrix, we introduce the following tensor

$$M_0 = -\frac{1}{2}I_2 + (\rho_0 \otimes y_0) + \frac{1}{2}\int_{\partial\omega} u_0^2 (y \otimes n)\,d\sigma + \int_{\partial\omega} u_0 (\chi \otimes n)\,d\sigma. \quad (2.23)$$

In the next proposition we show that function $\Lambda_0$ defined in (2.18) is differentiable at $\xi = 0$ with $\nabla\Lambda_0(0) = \rho_0$. Furthermore we give a polynomial estimate at third order for $\Lambda_0(\xi)$ as well as for the following "error" functional

$$E_\xi(v) := \int_\omega (1-\xi\cdot y)\,(|\nabla v|^2 - (\lambda_0 + \rho_0\cdot\xi)\,v^2)dy + \int_{\partial\omega}(1-\xi\cdot y)\,\gamma\,v^2\,d\sigma. \quad (2.24)$$

**Proposition 2.1.** *Let $\rho_0, M_0$ be defined by (2.21) and (2.23), respectively. Then, there exists constants $C > 0$ and $r_0 > 0$ such that:*

$$\left|\Lambda_0(\xi) - \left(\lambda_0 + \rho_0\cdot\xi + \frac{1}{2}M_0\,\xi\cdot\xi\right)\right| \leq C\,|\xi|^3 \qquad \text{whenever } |\xi| < r_0. \quad (2.25)$$

$$\left|E_\xi(u_0 + \chi_\xi) - \frac{1}{2}M_0\xi\cdot\xi\right| \leq C\,|\xi|^3 \qquad \text{whenever } |\xi| < r_0. \quad (2.26)$$

We notice that if, in the expressions (2.21) and (2.23), we substitute $u_0$ with the fundamental mode of the Dirichlet problem in $\Omega$, then we obtain $\rho_0 = 0$ and $M_0 = -\frac{1}{2}I_2$ which is nothing else but the result in [2] (Proposition 4.1).

The identification of the first and second order terms of $\Lambda_0(\xi)$ near $\xi = 0$ can be done directly by formal asymptotic expansion. However, in order to provide a rigorous proof, we will use an alternative formulae for matrix $M_0$, given in next lemma.

**Lemma 2.2.** *Let $\rho_0$ and $M_0$ be given by (2.21) and by (2.23), respectively. Then, the following equalities hold true for every $\xi \in \mathbb{R}^2$:*

$$\frac{1}{2}M_0\xi\cdot\xi = \int_\omega (\xi\cdot\nabla\chi_\xi)u_0 + (\xi\cdot\nabla u_0)(\xi\cdot y)u_0\,dy = \int_\omega (\xi\cdot\nabla u_0)\,\chi_\xi + (\rho_0\cdot\xi)(y_0\cdot\xi). \quad (2.27)$$

*Proof.* Since $\text{div}((\xi\cdot y)\xi) = \|\xi\|^2$ and $\|u_0\|_{L^2(\omega)} = 1$, by integrating by parts we obtain

$$\int_\omega (\xi\cdot\nabla|u_0|^2)(\xi\cdot y)\,dy = -\|\xi\|^2 + \int_{\partial\omega}(\xi\cdot n)(\xi\cdot y)\,|u_0|^2\,d\sigma$$

and, consequently,

$$\int_\omega (\xi\cdot\nabla u_0)(\xi\cdot y)u_0\,dy = -\frac{\|\xi\|^2}{2} + \frac{1}{2}\int_{\partial\omega}(\xi\cdot n)(\xi\cdot y)\,|u_0|^2\,d\sigma. \quad (2.28)$$

On the other hand, noticing that $\xi = \nabla(\xi\cdot y)$ and exploiting equations (2.20) and (2.22), we infer

$$\int_\omega (\xi \cdot \nabla \chi_\xi) \, u_0 \, dy - \int_\omega (\xi \cdot \nabla u_0) \, \chi_\xi \, dy \;=\; \int_\omega \xi \left( \nabla \chi_\xi \, u_0 - \nabla u_0 \, \chi_\xi \right) dy$$

$$= \; -\int_\omega (\xi \cdot y)(\xi \cdot \nabla u_0) \, u_0 \, dy + (\rho_0 \cdot \xi)(y_0 \cdot \xi)$$

$$= \; \frac{\|\xi\|^2}{2} - \frac{1}{2} \int_{\partial \omega} (\xi \cdot n)(\xi \cdot y) \, |u_0|^2 \, d\sigma + (\rho_0 \cdot \xi)(y_0 \cdot \xi)$$

where in the last line we exploit identity (2.28). Noticing that

$$\int_\omega (\xi \cdot \nabla \chi_\xi) \, u_0 \, dy + \int_\omega (\xi \cdot \nabla u_0) \, \chi_\xi \, dy = \int_{\partial \omega} (\xi \cdot n) \, \chi_\xi \, u_0 \, d\sigma,$$

we deduce that

$$\int_\omega (\xi \cdot \nabla \chi_\xi) \, u_0 \, dy = \frac{\|\xi\|^2}{4} - \frac{1}{4} \int_{\partial \omega} (\xi \cdot n)(\xi \cdot y) \, |u_0|^2 \, d\sigma + \frac{1}{2}(\rho_0 \cdot \xi)(y_0 \cdot \xi) + \frac{1}{2} \int_{\partial \omega} (\xi \cdot n) \, \chi_\xi \, u_0 \, d\sigma \quad (2.29)$$

$$\int_\omega (\xi \cdot \nabla u_0) \, \chi_\xi \, dy = -\frac{\|\xi\|^2}{4} + \frac{1}{4} \int_{\partial \omega} (\xi \cdot n)(\xi \cdot y) \, |u_0|^2 \, d\sigma - \frac{1}{2}(\rho_0 \cdot \xi)(y_0 \cdot \xi) - \frac{1}{2} \int_{\partial \omega} (\xi \cdot n) \, \chi_\xi \, u_0 \, d\sigma \quad (2.30)$$

Plugging (2.28), (2.29) and (2.30) in the second and third members of (2.27), it can be checked that both expressions agree with $\frac{1}{2} M_0 \xi \cdot \xi$, being $M_0$ given by (2.23).

$\square$

**Proof of Proposition 2.1**

We begin by proving (2.25). This is done in two steps.

**Step 1.** First we notice that the perturbed eigenvalue problem (2.17) is well posed provided $\xi$ is small enough. Indeed, if $1 - \xi \cdot y$ has a positive lower bound on $\omega$, then the operator

$$A_\xi : w \mapsto -\mathrm{div}\Big((1 - \xi \cdot y) \nabla w\Big), \quad D(A_\xi) = \left\{ w \in H^2(\omega) \; : \; \frac{\partial w}{\partial n} + \gamma w = 0, \text{ on } \partial \omega \right\}$$

has compact resolvent and is a positive self-adjoint operator, acting on $L^2(\omega)$ endowed with the scalar product $(u|v) = \int_\omega (1 - \xi \cdot y) \, u \, \bar{v} \, dy$. As a consequence of Krein Rutman's Theorem, the first eigenvalue $\Lambda_0(\xi)$ is simple and the second eigenvalue $\Lambda_1(\xi)$ is such that $\Lambda_1(\xi) > \Lambda_0(\xi)$. In fact there exits $r_0 > 0$ and $\kappa > 0$ such that

$$\Lambda_1(\xi) - \Lambda_0(\xi) \;\geq\; \kappa \quad \text{whenever } |\xi| < r_0, \quad (2.31)$$

which follows from the continuity of functions $\Lambda_0$ and $\Lambda_1$ in a neighborhood of $\xi = 0$. This fact can be established by using the strong continuity with respect to $\xi$ of the resolvent operator or, directly, by passing to the limit in the variational characterization of $\Lambda_0(\xi_n)$ and $\Lambda_1(\xi_n)$ on a sequence $\xi_n \to \xi$, with the help of the compact embedding $H^1(\Omega) \subset L^2(\Omega)$.

Let us set $P(\xi) := \lambda_0 + \rho_0 \cdot \xi + \frac{1}{2} M_0 \xi \cdot \xi$. By the continuity of $|\Lambda_0(\xi) - P(\xi)|$, we need only to prove that

$$\limsup_{\xi \to 0, \, \xi \neq 0} \left\{ \frac{|\Lambda_0(\xi) - P(\xi)|}{|\xi|^3} \right\} \;<\; +\infty.$$

To that aim we substitute $\xi$ by $\varepsilon \xi$, where $\varepsilon \to 0$ and $|\xi| = 1$, and show that $|\Lambda_0(\varepsilon \xi) - P(\varepsilon \xi)| \leq C \varepsilon^3$, for a suitable constant $C$ (independent of $\varepsilon$ and of the unit vector $\xi$).

Exploiting (2.31), we may apply the assertion i) of Lemma 5.1 to the operator $A_{\varepsilon \xi}$ defined in the Hilbert space $H_\varepsilon = L^2(\omega)$ endowed with the scalar product $(u|v)_\varepsilon := \int_\omega (1 - \varepsilon \xi \cdot y) \, u \, \bar{v} \, dy$.

As we have $(1-C\varepsilon)\|u\|_{L^2(\omega)} \leq (u|u)_\varepsilon \leq (1+C\varepsilon)\|u\|_{L^2(\omega)}$, we eventually conclude that (2.25) holds true provided we show the existence of a sequence of quasi eigenvector $\{w_\varepsilon\}$ such that:

$$\|A_{\varepsilon\xi}w_\varepsilon - P(\varepsilon\xi)w_\varepsilon\|_{L^2(\omega)} \leq C\varepsilon^3 \|w_\varepsilon\|_{L^2(\omega)}. \tag{2.32}$$

**Step 2.** We prove (2.32). In what follows we will suppose that $\xi$ is fixed and, in order to simplify the computations, we denote $\lambda_1 := \rho_0 \cdot \xi$ (see (2.21)) and $u_1 := \chi_\xi$, so that problem (2.22) reads

$$\begin{cases} -\Delta u_1 - \lambda_0\, u_1 = -\xi \cdot \nabla u_0 + \lambda_1 u_0, & \text{in } \omega, \\ \dfrac{\partial u_1}{\partial n} + \gamma u_1 = 0, & \text{on } \partial\omega, \\ \int_\omega u_1 u_0\, dy = 0. \end{cases} \tag{2.33}$$

Setting $\lambda_2 := \frac{1}{2}M_0 \xi \cdot \xi$, we have $P(\varepsilon\xi) = \lambda_0 + \varepsilon\lambda_1 + \varepsilon^2\lambda_2$. Let us consider

$$w_\varepsilon := u_0 + \varepsilon u_1 + \varepsilon^2 u_2, \tag{2.34}$$

where $u_2$ is the unique solution of

$$\begin{cases} -\Delta u_2 - \lambda_0\, u_2 = -\xi \cdot \nabla u_1 - (\xi \cdot \nabla u_0)(\xi \cdot y) + \lambda_1 u_1 + \lambda_2 u_0, & \text{in } \omega, \\ \frac{\partial u_2}{\partial n} + \gamma u_2 = 0, & \text{on } \partial\omega, \\ \int_\omega u_2 u_0\, dy = 0. \end{cases} \tag{2.35}$$

The existence of $u_2$ follows from the Fredholm orthogonality condition

$$\lambda_2 = \int_\omega [(\xi \cdot \nabla u_1)u_0 + (\xi \cdot \nabla u_0)(\xi \cdot y)u_0]\, dy,$$

which by (2.27) is satisfied precisely for $\lambda_2 := \frac{1}{2}M_0\xi \cdot \xi$. On the other hand $w_\varepsilon$ given by (2.34) satisfies the prescribed Robin condition and therefore belongs to the domain of $A_{\varepsilon\xi}$.
Now we compute $A_{\varepsilon\xi}(w_\varepsilon) - P(\varepsilon\xi)$ gathering power like terms in $\varepsilon$ and using (2.20), (2.33), (2.34) and (2.35):

$$\begin{aligned}
-\mathrm{div}\Big([1-(\varepsilon\xi \cdot y)]\nabla w_\varepsilon\Big) - P(\varepsilon\xi)[1-(\varepsilon\xi \cdot y)]w_\varepsilon =\ & -\varepsilon^3[(\xi \cdot y)\Delta u_2 + \xi \cdot \nabla u_2] \\
& -\varepsilon^3[(\xi \cdot y)(\lambda_0 u_2 + \lambda_1 u_1 + \lambda_2 u_0) - \lambda_1 u_2 - \lambda_2 u_1] \\
& -\varepsilon^4[(\xi \cdot y)(\lambda_1 u_2 + \lambda_2 u_1) - \lambda_2 u_2] \\
& -\varepsilon^5[(\xi \cdot y)\lambda_2 u_2].
\end{aligned}$$

In view of the continuous polynomial dependence of $u_1, u_2, \lambda_1$ and $\lambda_2$ with respect to $\xi$, and since $\|w_\varepsilon\|_{L^2} \to 1$, we can therefore find a constant $C > 0$ independent of $\varepsilon$ such that (2.32) holds true. This completes the proof of (2.25).

**Proof of (2.26):** In view of (2.20), we have that $-\mathrm{div}((1-\xi \cdot y)\nabla u_0) = \lambda_0(1-\xi \cdot y)u_0 + \xi \cdot \nabla u_0$. By integration by parts, we deduce that for every $\psi \in H^1(\omega)$ it holds

$$\int_\omega (1-\xi \cdot y)(\nabla u_0 \nabla \psi - \lambda_0\, u_0\, \psi) + \int_{\partial\omega}(1-\xi \cdot y)\gamma\, u_0\, \psi = \int_\omega (\xi \cdot \nabla u_0) \tag{2.36}$$

In particular, for $\psi = u_0$, taking into account that $\int_\omega u_0^2 = 1$ and (2.21), we obtain:

$$E_\xi(u_0) = -\int_\omega \rho_0 \cdot \xi(1-\xi \cdot y)u_0^2 + \int_\omega (\xi \cdot \nabla u_0)\, u_0 = (\xi \cdot \rho_0)(\xi \cdot y_0) - \xi \cdot \rho_0 + \frac{1}{2}\xi \cdot \int_\omega \nabla(u_0^2) = (\xi \cdot \rho_0)(\xi \cdot y_0).$$

Taking now $\psi = \chi_\xi$ in (2.36) and recalling that $\int_\omega u_0 \chi_\xi = 0$, we get

$$\begin{aligned} E_\xi(u_0 + \chi_\xi) &= E_\xi(u_0) + E_\xi(\chi_\xi) + 2\left[\int_\omega (1-\xi\cdot y)\left(\nabla u_0 \nabla \chi_\xi - (\lambda_0 + \xi\cdot\rho_0)\, u_0\, \chi_\xi\right) + \int_{\partial\omega}(1-\xi\cdot y)\,\gamma\, u_0\, \chi_\xi\right] \\ &= (\xi\cdot\rho_0)(\xi\cdot y_0) + E_\xi(\chi_\xi) + 2\int_\omega (\xi\cdot\nabla u_0)\,\chi_\xi + 2\,\xi\cdot\rho_0 \int_\omega (\xi\cdot y) u_0\, \chi_\xi\,. \end{aligned} \quad (2.37)$$

On the other hand, by (2.22) we have

$$-\mathrm{div}\Big((1-\xi\cdot y)\nabla\chi_\xi\Big) = (1-\xi\cdot y)(\lambda_0 \chi_\xi - \xi\cdot\nabla u_0 + \xi\cdot\rho_0\, u_0) + \xi\cdot\nabla\chi_\xi,$$

from which follows, by multiplying by $\chi_\xi$ and integrating by parts,

$$E_\xi(\chi_\xi) = -\int_\omega (1-\xi\cdot y)\left[(\xi\cdot\rho_0)\,(|\chi_\xi|^2 - u_0\,\chi_\xi) + \xi\cdot\nabla u_0\,\chi_\xi\right] + \int_\omega \xi\cdot\nabla\chi_\xi\,\chi_\xi\,.$$

Then, recalling that $\int_\omega u_0\chi_\xi = 0$, we rewrite (2.37) as follows

$$E_\xi(u_0 + \chi_\xi) = (\xi\cdot\rho_0)(\xi\cdot y_0) + \int_\omega (\xi\cdot\nabla u_0)\,\chi_\xi + R(\xi)\,,$$

where the reminder $R(\xi)$ is a sum of terms of power order greater than 3 with respect to $\xi$:

$$R(\xi) = \int_\omega (\xi\cdot y)\,\xi\cdot\nabla u_0\,\chi_\xi - \int_\omega (1-\xi\cdot y)(\xi\cdot\rho_0)\,|\chi_\xi|^2 + (\xi\cdot\rho_0)\int_\omega (\xi\cdot y)\,u_0\,\chi_\xi + \int_\omega (\xi\cdot\nabla\chi_\xi)\,\chi_\xi\,.$$

Therefore, taking into account the second equality in (2.27), we conclude that, for $|\xi|$ sufficiently small,

$$\left| E_\xi(u_0 + \chi_\xi) - \frac{1}{2} M_0 \xi\cdot\xi \right| = |R(\xi)| \leq C\,|\xi|^3\,.$$

$\square$

## 3. The symmetric case

In this section we will assume that the solution $u_0$ of (1.1) satisfies the following balance relation:

$$\rho_0 = \int_{\partial\omega} u_0^2\, n\, d\sigma = 0. \quad (3.1)$$

This condition is necessary in order that, for small values of $\varepsilon$, the lower energy modes propagate along the $x_3$ direction. Otherwise, as we will discover in the next section, the fundamental mode will localize. Let us notice that condition (3.1) involves only the geometry of $\omega$ and the function $\gamma \in L^\infty(\partial\omega)$ associated with the Robin condition. In particular, if $\gamma$ is constant, it can be checked that it is fulfilled if $\omega$ has one axis of symmetry.

We will assume further that the curvature $k(s)$, the torsion $\tau(s)$ and the angular parameter $\alpha(s)$ have the following regularity

$$k \in L^\infty(0,L) \quad,\quad \tau \in W^{1,\infty}(0,L) \quad,\quad \alpha \in W^{2,\infty}(0,L)\,. \quad (3.2)$$

Then, recalling (2.8) (in particular $\tilde\tau = \tau + \alpha'$) and (2.23), we set

$$q(s) := \frac{1}{2} M_0 \xi(s)\cdot\xi(s) + C_1\,\tilde\tau(s)^2 - C_2 \tilde\tau'(s) \quad,\quad \xi(s) = k(s) z_{\alpha(s)}\,, \quad (3.3)$$

where constants $C_1, C_2$ (depending on $\omega$ and $u_0$) are defined as follows:

$$C_1 := \int_\omega |\nabla u_0 \cdot Ry|^2 \, dy + \frac{1}{2} \int_{\partial\omega} \gamma \, u_0^2 \, (y \cdot \dot{y})^2 \, d\sigma \quad , \quad C_2 := \int_\omega u_0 (\nabla u_0 \cdot Ry) \, dy. \tag{3.4}$$

The scalar function $q(s)$ appearing in (3.3) will play the role of an effective potential in the limit problem which rules the $x_3$-propagation of lower order modes. More precisely, let us introduce the following Sturm-Liouville problem

$$\begin{cases} -w'' + q(s) \, w = \mu \, w, & w \in H^2(0, L), \\ -w'(0) + \left( \tilde{\gamma}_0 - C_2 \tilde{\tau}(0) \right) w(0) = 0, \\ w'(L) + \left( \tilde{\gamma}_L + C_2 \tilde{\tau}(L) \right) w(L) = 0, \end{cases} \tag{3.5}$$

where we have set

$$\tilde{\gamma}_0 := \int_\omega \gamma_0 \, u_0^2 \, dy \quad , \quad \tilde{\gamma}_1 := \int_\omega \gamma_1 \, u_0^2 \, dy. \tag{3.6}$$

Then, the main result of this section states the convergence of the family of spectral problems (1.1) in the symmetric case.

**Theorem 3.1.** *Assume that (3.1) and (3.2) hold. Then the eigenvalues $\lambda_0^\varepsilon < \lambda_1^\varepsilon \leq \cdots \leq \lambda_i^\varepsilon \leq \cdots$ of the spectral problem (1.1) satisfy for each $i \in \mathbb{N}$*

$$\lambda_i^\varepsilon = \frac{\lambda_0}{\varepsilon^2} + \mu_i^\varepsilon \quad , \quad \mu_i^\varepsilon \to \mu_i, \tag{3.7}$$

*where $\mu_i$ $(i \in \mathbb{N})$ are the eigenvalues of (3.5). Furthermore, if $u_i^\varepsilon$ is a normalized eigenvector for problem (1.1) associated with $\lambda_i^\varepsilon$, then, up to a subsequence, $v_i^\varepsilon(s, y) = u_i^\varepsilon(\psi_\varepsilon(s, y))$ converges strongly in $L^2(Q_L)$ to $v_i(s, y) = w_i(s) u_0(y)$ where $w_i$ is a normalized eigenvector of problem (3.5) associated with $\mu_i$. Conversely, any such $v_i$ is the limit of a sequence $u_i^\varepsilon \circ \psi_\varepsilon$ where $u_i^\varepsilon$ is an eigenvector of (1.1) associated with $\lambda_i^\varepsilon$.*

**Remark 3.2.** The result above is quite similar to the main result of [2]. Only changes the structure of the effective potential $q(s)$. In particular, the influence of the curvature $k(s)$ is taken into account through the function $M_0 \xi(s) \cdot \xi(s)$ where $M_0$ is not a priori a scalar tensor as it was in [2]. Notice that if we formally substitute the Robin condition on the lateral part of the tube by a Dirichlet one (that is $\gamma = +\infty$), we get $M_0 = -\frac{1}{2} I_2$ (independent of the shape of $\omega$) and we recover the effective potential obtained in [2].

The key argument in order to prove Theorem 3.1 consists in establishing the $\Gamma$-convergence of suitable quadratic energies defined on $H^1(Q_L)$ and to apply to them the general statement of Proposition (5.2) (see Appendix). We observe that, thanks to (3.1), applying (2.25) yields

$$\frac{\Lambda_0(\varepsilon \xi(s)) - \lambda_0}{\varepsilon^2} \to \frac{1}{2} M_0 \, \xi(s) \cdot \xi(s) \quad \text{uniformly on } [0, L]. \tag{3.8}$$

In view of (3.8), these functionals are obtained, up to multiplicative factor $\varepsilon^2$, by shifting the initial energy $\tilde{F}_\varepsilon$. More precisely, we introduce $\tilde{G}_\varepsilon : L^2(Q_L) \to \overline{\mathbb{R}}$ defined by

$$\tilde{G}_\varepsilon(v) := \begin{cases} \frac{1}{\varepsilon^2} \tilde{F}_\varepsilon(v) - \frac{1}{\varepsilon^2} \int_0^L \int_\omega \beta_\varepsilon \, \lambda_0 \, v^2 dy, & \text{if } v \in H^1(Q_L), \\ +\infty & \text{otherwise.} \end{cases} \tag{3.9}$$

In connection with spectral problem (3.5), we consider $\tilde{G} : L^2(Q_L) \to \overline{\mathbb{R}}$ defined as follows

$$\tilde{G}(v) := \begin{cases} \tilde{G}_0(w) & \text{if } v(s, y) = w(s) \, u_0(y), \ w \in H^1(0, L), \\ +\infty & \text{otherwise,} \end{cases} \tag{3.10}$$

where

$$\tilde{G}_0(w) := \int_0^L \left( |w'|^2 + \left[ C_1\tilde{\tau}^2 + \frac{1}{2}M_0\xi \cdot \xi \right]w^2 + 2C_2\tilde{\tau}(s)w'w \right) ds + \tilde{\gamma}_0\, w(0)^2 + \tilde{\gamma}_L\, w(L)^2. \quad (3.11)$$

**Proposition 3.3.** *Under the hypotheses of Theorem 3.1, $\tilde{G}_\varepsilon$ $\Gamma$ - converges in $L^2(Q)$ to $\tilde{G}$ given by (3.10) and (3.11). Moreover, the family of functionals $\tilde{G}_\varepsilon$ satisfies all conditions i), ii) and iii) of Proposition 5.2.*

*Proof.* We proceed in three steps: in Step 1 we prove that $\{\tilde{G}_\varepsilon\}$ satisfies the hypothesis $i)$ and $ii)$. Then we split the proof of $iii)$ into two steps: in Step 2 we prove the lower bound inequality for the $\Gamma$-convergence, and in Setp 3 we establish the existence of a sequence realizing the lower bound.

**Step 1.** Recalling the definition of $\tilde{G}_\varepsilon$ (see (3.9) and (2.14)) we have

$$\begin{aligned}
\tilde{G}_\varepsilon(v) &= \int_0^L \int_\omega \frac{1}{\beta_\varepsilon}\left|v' + (\nabla_y v \cdot R\, y)\tilde{\tau}\right|^2 dy\, ds \\
&+ \frac{1}{\varepsilon^2}\int_0^L \left[ \int_\omega \beta_\varepsilon(|\nabla_y v|^2 - \lambda_0|v|^2)dy + \int_{\partial\omega}\beta_\varepsilon\, \gamma\, |v|^2 d\sigma \right] ds \\
&+ \int_0^L \frac{\tilde{\tau}^2}{2}\left( \int_{\partial\omega} \gamma\, |v|^2(y \cdot \dot{y})^2\, d\sigma \right) ds + \varepsilon \int_0^L \int_{\partial\omega} \gamma\, |v|^2\, r_\varepsilon\, d\sigma\, ds \\
&+ \int_\omega \left( \gamma_0\, |v(0,y)|^2 + \gamma_L\, |v(L,y)|^2 \right) dy,
\end{aligned} \quad (3.12)$$

where $r_\varepsilon(s,y)$ is uniformly bounded in $(s,y)$, for $\varepsilon$ small enough (see (2.10) and (2.11)).

Since $\gamma, \gamma_0, \gamma_L$ and $\dfrac{\tilde{\tau}^2}{2}(y \cdot \dot{y})^2 + \varepsilon r_\varepsilon$ are non negative (see (2.11)), from (3.12) we deduce

$$\begin{aligned}
\tilde{G}_\varepsilon(v) &\geq \int_0^L \int_\omega \frac{1}{\beta_\varepsilon}\left|v' + (\nabla_y v \cdot R\, y)\tilde{\tau}\right|^2 dy\, ds \\
&+ \frac{1}{\varepsilon^2}\int_0^L \left[ \int_\omega \beta_\varepsilon(|\nabla_y v|^2 - \lambda_0|v|^2)dy + \int_{\partial\omega}\beta_\varepsilon\, \gamma\, |v|^2 d\sigma \right] ds,
\end{aligned} \quad (3.13)$$

and also, using the definition of $\Lambda_0(\varepsilon\xi(s))$ (see (2.18)),

$$\begin{aligned}
\tilde{G}_\varepsilon(v) &\geq \int_0^L \int_\omega \left[ \frac{1}{\beta_\varepsilon}\left|v' + (\nabla_y v \cdot R\, y)\tilde{\tau}\right|^2 + \beta_\varepsilon \frac{\Lambda_0(\varepsilon\xi(s)) - \lambda_0}{\varepsilon^2}|v|^2 \right] dy\, ds \\
&+ \int_0^L \frac{\tilde{\tau}^2}{2}\left( \int_{\partial\omega} \gamma\, |v|^2(y \cdot \dot{y})^2\, d\sigma \right) ds + \varepsilon \int_0^L \int_{\partial\omega} \gamma\, |v|^2\, r_\varepsilon\, d\sigma\, ds \\
&+ \int_\omega \left( \gamma_0\, |v(0,y)|^2 + \gamma_L\, |v(L,y)|^2 \right) dy \\
&\geq \int_0^L \int_\omega \left[ \frac{1}{\beta_\varepsilon}\left|v' + (\nabla_y v \cdot R\, y)\tilde{\tau}\right|^2 + \beta_\varepsilon \frac{\Lambda_0(\varepsilon\xi(s)) - \lambda_0}{\varepsilon^2}|v|^2 \right] dy\, ds.
\end{aligned} \quad (3.14)$$

Since $\beta_\varepsilon$ converges uniformly to 1, in view of (3.8) and (3.14), for $\varepsilon$ small enough we can find $c_0$ such that condition $i)$ is satisfied.

Consider now a sequence $\{v_\varepsilon\}$ bounded in $L^2(Q_L)$, such that $\tilde{G}_\varepsilon(v_\varepsilon)$ is also uniformly bounded. Then, first from (3.13) and (3.8), and then from (3.14), we will obtain, for some $M$ and $N$ independent of $\varepsilon$,

$$\int_{Q_L} \left|v'_\varepsilon + (\nabla_y v_\varepsilon \cdot R\, y)\, \tilde{\tau}\right|^2 \leq M, \quad \int_{Q_L} |\nabla_y v_\varepsilon|^2 \leq N. \tag{3.15}$$

From (3.15), we infer that the sequence $\{Dv_\varepsilon\}$, where $Dv_\varepsilon = (v'_\varepsilon, \nabla_y v_\varepsilon)$, is bounded in $[L^2(Q_L)]^3$. Thus $\{v_\varepsilon\}$ is bounded in $H^1(Q_L)$ and strongly relatively compact in $L^2(Q_L)$ by Rellich-Kondrachov Theorem.

**Step 2.** Let $\{v_\varepsilon\}$ be a sequence such that $v_\varepsilon \to v$ in $L^2(Q_L)$. Up to a subsequence we may assume that $\liminf_{\varepsilon \to 0} G_\varepsilon(v_\varepsilon) = \lim_{\varepsilon \to 0} \tilde{G}_\varepsilon(v_\varepsilon) < +\infty$. Then, as proved in Step 1, the sequence is bounded in $H^1(Q_L)$ and inequalities (3.15) apply. Therefore, $v$ belongs to $H^1(Q_L)$ and $v'_\varepsilon \rightharpoonup v', \nabla_y v_\varepsilon \rightharpoonup \nabla_y v$ weakly in $L^2(Q_L)$. In particular, as $R\, y\, , \tilde{\tau} \in L^\infty(Q_L)$, we obtain:

$$v'_\varepsilon + (\nabla_y v_\varepsilon \cdot R\, y)\tilde{\tau} \quad \rightharpoonup \quad v' + (\nabla_y v \cdot R\, y)\, \tilde{\tau}\ .$$

Futhermore, from (3.14) and the uniform convergence (3.8) we deduce that

$$\liminf_{\varepsilon \to 0} \tilde{G}_\varepsilon(v_\varepsilon) \geq \int_{Q_L} \left\{\left|v' + (\nabla_y v \cdot R\, y)\, \tilde{\tau}\right|^2 + \frac{1}{2}\Big(M_0\, \xi(s) \cdot \xi(s)\Big)|v|^2\right\}\, dy\, ds$$
$$+ \int_0^L \frac{\tilde{\tau}^2}{2}\left(\int_{\partial \omega} \gamma\, |v|^2 (y \cdot \dot{y})^2\, d\sigma\right)\, ds + \int_\omega \left(\gamma_0\, |v(0,y)|^2 + \gamma_L\, |v(L,y)|^2\right)\, dy. \tag{3.16}$$

On the other hand, from (3.13) and since $\tilde{G}_\varepsilon(v_\varepsilon)$ is uniformly bounded, one has that

$$0 \geq \liminf_{\varepsilon \to 0} \int_0^L \left(\int_\omega \beta_\varepsilon(|\nabla_y v_\varepsilon|^2 - \lambda_0|v_\varepsilon|^2)dy + \int_{\partial\omega} \beta_\varepsilon\, \gamma\, |v_\varepsilon|^2 d\sigma\right)\, ds.$$

But

$$\liminf_{\varepsilon \to 0} \int_0^L \int_\omega \beta_\varepsilon(|\nabla_y v_\varepsilon|^2 - \lambda_0|v_\varepsilon|^2)\, dy\, ds + \int_0^L \int_{\partial\omega} \beta_\varepsilon\, \gamma\, |v_\varepsilon|^2\, d\sigma\, ds \geq$$
$$\int_0^L \int_\omega |(\nabla_y v|^2 - \lambda_0|v|^2)dyds + \int_0^L \int_{\partial\omega} \gamma\, |v|^2 d\sigma ds \geq 0,$$

by the definition of $\lambda_0$. Therefore, for a.e. $s \in (0, L)$,

$$\int_\omega |(\nabla_y v|^2 - \lambda_0|v|^2)dy + \int_{\partial\omega} \gamma\, |v|^2 d\sigma = 0$$

and $v(s, \cdot)$, as an eigenvector associated with $\lambda_0$, is proportional to the ground state $u_0$. We deduce that $v$ can be written in the form $v(s, y) = w(s)\, u_0(y)$ with $w \in H^1(0, L)$ (since $v \in H^1(Q_L)$). We plug this expression of $v$ into (3.16) to conclude that $\liminf_{\varepsilon \to 0} \tilde{G}_\varepsilon(v_\varepsilon) \geq \tilde{G}(v)$ where $\tilde{G}(v) = \tilde{G}_0(w)$. This achieves the proof of the lower bound for the Γ-convergence.

**Step 3.** Let $v \in L^2(Q_L)$. We have to show the existence of a sequence $\{v_\varepsilon\}$ such that $v_\varepsilon \to v$ and $\lim_{\varepsilon \to 0} \tilde{G}_\varepsilon(v_\varepsilon) = \tilde{G}(v)$. We may assume that $\tilde{G}(v) < +\infty$ so that we can write $v(s, y) = w(s)u_0(y)$ for a suitable element $w \in H^1(0, L)$. We consider $v_\varepsilon$ defined by $v_\varepsilon = w(s)[u_0(y) + \varepsilon\varphi(s, y)]$ where $\varphi \in H^1(Q_L)$ is given by $\varphi(s, y) = \chi_{\xi(s)}(y)$, $\chi_{\xi(s)}$ being the solution, for each $s$, of problem (2.22), for $\xi = \xi(s)$. Clearly $v_\varepsilon \to v$ strongly in $H^1(Q_L)$ and, as $\beta_\varepsilon$ is uniformly close to 1, we have

$$\lim_{\varepsilon \to 0} \left[ \int_{Q_L} \frac{1}{\beta_\varepsilon} \Big| v'_\varepsilon + \nabla_y v_\varepsilon \cdot R \, y \, \tilde{\tau} \Big|^2 ds \, dy + \int_0^L \frac{\tilde{\tau}^2}{2} \Big( \int_{\partial \omega} \gamma |v_\varepsilon|^2 (y \cdot \dot{y})^2 d\sigma \Big) ds \right]$$

$$= \int_{Q_L} \Big| v' + \nabla_y v \cdot R \, y \, \tilde{\tau} \Big|^2 ds \, dy + \int_0^L \frac{\tilde{\tau}^2}{2} \Big( \int_{\partial \omega} \gamma \, |v|^2 (y \cdot \dot{y})^2 \, d\sigma \Big) ds \qquad (3.17)$$

$$= \int_0^L |w'|^2 + C_1 \tilde{\tau}(s))^2 |w|^2 + C_2 \, w' \, w \, \tilde{\tau}(s) \, ds$$

and

$$\lim_{\varepsilon \to 0} \int_\omega \Big( \gamma_0 \, |v_\varepsilon(0,y)|^2 + \gamma_L \, |v_\varepsilon(L,y)|^2 \Big) \, dy = \tilde{\gamma}_0 \, |w(0)|^2 + \tilde{\gamma}_L \, |w(L)|^2. \qquad (3.18)$$

On the other hand, since $v_\varepsilon = w(s) \, (u_0 + \chi_{\varepsilon\xi})$, replacing $\beta_\varepsilon$ by $[1 - \varepsilon(\xi \cdot y)]$ and using assertion ii) of Proposition 2.1 with $\rho_0 = 0$, we obtain

$$\frac{1}{\varepsilon^2} \Big[ \int_\omega \beta_\varepsilon(s,y) \Big( |\nabla_y v_\varepsilon|^2 - \lambda_0 |v_\varepsilon|^2 \Big) dy + \int_{\partial\omega} \beta_\varepsilon(s,y) \, \gamma \, |v_\varepsilon|^2 \, d\sigma \Big] =$$
$$= \frac{1}{\varepsilon^2} |w(s)|^2 E_{\varepsilon\xi(s)} \big( u_0 + \chi_{\varepsilon\xi(s)} \big) = |w(s)|^2 \, \frac{1}{2} \, M_0 \, \xi(s) \cdot \xi(s) + \rho_\varepsilon(s), \qquad (3.19)$$

where $\lim_{\varepsilon \to 0} \rho_\varepsilon(s) = 0$, uniformly in $[0, L]$.

Passing to the limit in $\tilde{G}_\varepsilon(v_\varepsilon)$ as $\varepsilon \to 0$ and taking into account (3.17), (3.18) and (3.19) integrated with repect to $s$, we are led to

$$\lim_{\varepsilon \to 0} \tilde{G}_\varepsilon(v_\varepsilon) = \int_0^L \left[ |w'|^2 + \Big( C_1 \tilde{\tau}^2 + \frac{1}{2} M_0 \xi \cdot \xi \Big) |w|^2 + 2 C_2 \tilde{\tau}(s) w' w \right] ds + \tilde{\gamma}_0 w(0)^2 + \tilde{\gamma}_L w(L)^2,$$

which completes the proof. $\square$

**Proof of Theorem 3.1** By Proposition 3.3, $\tilde{G}$ given by (3.11) is nothing else but the $\Gamma$-limit in $L^2(Q_L)$ of $(\tilde{G}_\varepsilon)$ as $\varepsilon \to 0$. It is a lower semicontinuous and quadratic functional from $L^2(Q_L)$ into $(-\infty, +\infty]$ (in the sense of ([5], Theorem 11.10). By (3.11) its domain of finiteness $D(\tilde{G}) = \{w(s) \, u_0(y) : w \in H^1(0,L)\}$ can be identified with $H^1(0,L)$ and we have

$$\tilde{G}(w(s) u_0(y)) = \tilde{G}_0(w) = a_0(w,w) \, ,$$

where $a_0$ is the continuous coercive bilinear symmetric form on $H^1(0,L)$ deduced from the right hand side of (3.11). After integration by parts and recalling the definition of $q$ in (3.3), we observe that for every smooth test function $\varphi$, there holds

$$a_0(w, \varphi) = \int_0^L (w' \varphi' + q \, w \varphi) \, ds + C_2 \Big( [\tilde{\tau} w \varphi](L) - [\tilde{\tau} w \varphi](0) \Big) + \tilde{\gamma}_0 w(0) \varphi(0) + \tilde{\gamma}_L w(L) \varphi(L).$$

Therefore, the self-adjoint operator associated with $a_0$ is the (compact resolvent) operator $A_0 : L^2(Q_L) \to L^2(Q_L)$ whose domain $D(A_0)$ consists of all elements $w \in H^2(0,L)$ which satisfy the boundary conditions appearing in (3.5) and such that $A_0 w = -w'' + q(s) \, w$ for all $w \in D(A_0)$. Then, Theorem 3.1 follows by applying Proposition 5.2 to the sequence $\{\tilde{G}_\varepsilon\}$, which by Proposition 3.3 satisfies all the required conditions. $\square$

## 4. Non symmetric case and localization

In this section we consider a geometry $\omega$ and a Robin factor $\gamma(s)$ for which the balance condition (3.1) is not satisfied, that is $\rho_0 = \dfrac{1}{2} \displaystyle\int_{\partial \omega} u_0^2 \, n \, d\sigma$ is a non zero vector.

It turns out that localization occurs at the minimum points of the following scalar product

$$\varphi(s) := \rho_0 \cdot \xi(s) \qquad (\text{recall } \xi(s) = k(s) z_\alpha(s)) \,. \tag{4.1}$$

We will assume that the function $\varphi$ is of class $\mathcal{C}^2([0, L])$ and that it admits a *unique global minimizer* at $s_0 \in (0, L)$:

$$\mu_0 := \varphi(s_0) < \varphi(s) \quad \text{for all } s \neq s_0 \text{ and } \quad \varphi''(s_0) > 0. \tag{4.2}$$

In particular $\varphi'(s_0) = 0$ and the function $\dfrac{\varphi(s) - \varphi(s_0)}{|s - s_0|^2}$ extended by prescribing the value $\tfrac{1}{2}\varphi''(s_0)$ at $s = s_0$ is positive continuous on the whole interval $[0, L]$. Thus, there exists $\eta_0 > 0$ such that

$$\eta_0 \, |s - s_0|^2 \;\leq\; \varphi(s) - \mu_0 \;\leq\; \frac{1}{\eta_0}|s - s_0|^2 \qquad \text{for every } s \in [0, L] \,. \tag{4.3}$$

We are going to show that localization occurs in the vicinity of $r(s_0)$. The concentrating behavior of eigenvectors turns out to be described after a suitable blow-up by the lower level eigenfunctions of the classical 1D-quantum harmonic oscillator. More precisely let

$$\nu_0 \;:=\; \sqrt{\tfrac{1}{2} \, (\rho_0 \cdot \xi''(s_0))} \tag{4.4}$$

and consider the spectral problem

$$- \hat{w}'' + \nu_0^2 \, t^2 \, \hat{w} = \nu \, \hat{w}, \quad \hat{w} \in H^2(\mathbb{R}) \cap L^2(\mathbb{R}; t^2 dt), \tag{4.5}$$

(being $L^2(\mathbb{R}; t^2 dt)$ the subspace of functions $\hat{w} \in L^2(\mathbb{R})$ such that $\int_{\mathbb{R}} t^2 \, |\hat{w}|^2 \, dt < \infty$).

We recall (se for instance [4], & 2, Prop 25-26]) that we may associate with (4.5) a positive self-adjoint operator in $L^2(\mathbb{R})$ with compact resolvent and whose eigenvalues are all simple and given by

$$\nu_i \;=\; \nu_0 \, (1 + 2i) \;, \quad i \in \mathbb{N}. \tag{4.6}$$

Moreover, there exists an orthonormal basis of eigenfunctions in terms of Hermite polynomials as follows:

$$\hat{w}_i(t) = \frac{1}{[2^i \, i! \sqrt{\pi}]^{1/2}} \, \sqrt[4]{\nu_0} \; e^{-\nu_0 \frac{t^2}{2}} \, H_i(t \sqrt{\nu_0}) \;, \quad H_i(t) = (-1)^i \, e^{\frac{t^2}{2}} \frac{d^i}{dt^i}(e^{-\frac{t^2}{2}}) \tag{4.7}$$

In particular, the normalized fundamental mode $\hat{w}_0(t) := \sqrt[4]{\tfrac{\nu_0}{\pi}} \; e^{-\nu_0 \frac{t^2}{2}}$ satisfies

$$\int_{\mathbb{R}} |\hat{w}_0|^2 = 1 \;, \quad \int_{\mathbb{R}} |\hat{w}_0'|^2 = \frac{\nu_0}{2} \;, \quad \int_{\mathbb{R}} t^2 |\hat{w}_0|^2 = \frac{1}{2\nu_0} \;, \quad \int_{\mathbb{R}} t^4 |\hat{w}_0|^2 = \frac{3}{4\nu_0^2}. \tag{4.8}$$

Our second main result reads as follows

**Theorem 4.1.** *Assume that $\xi(s)$ given in (4.1) belongs to $\mathcal{C}^2([0, L])$ and that (4.2) is satisfied. Let $\nu_i$ be defined by (4.6). Then the eigenvalues $\lambda_0^\varepsilon < \lambda_1^\varepsilon \leq \cdots \leq \lambda_i^\varepsilon \leq \cdots$ of the spectral problem (1.1) can be expanded as follows*

$$\lambda_i^\varepsilon \;=\; \frac{\lambda_0}{\varepsilon^2} + \frac{\mu_0}{\varepsilon} + \frac{\nu_i^\varepsilon}{\sqrt{\varepsilon}} \quad \text{where, for each } i \in \mathbb{N}, \quad \lim_{\varepsilon \to 0} \nu_i^\varepsilon = \nu_i \,. \tag{4.9}$$

*Furthermore, if $u_i^\varepsilon$ is a normalized eigenvector for problem (1.1) associated with $\lambda_i^\varepsilon$, then, up to a subsequence, $\hat{v}_i^\varepsilon(t,y) = \varepsilon^{1/8} u_i^\varepsilon(\psi_\varepsilon(s_0 + \varepsilon^{1/4}t, y))$ converges strongly in $L^2(\mathbb{R} \times \omega)$ to $\hat{v}_i(t,y) = \pm \hat{w}_i(t) u_0(y)$, being $\hat{w}_i$ given by (4.7). Conversely, any such $\hat{v}_i$ is the limit of a sequence of eigenvectors $u_i^\varepsilon$ for (1.1) associated with $\lambda_i^\varepsilon$.*

Since the eigenvalues $\nu_i$ are simple, we infer from previous theorem that, for all $i$, the spectral distance $\lambda_{i+1}^\varepsilon - \lambda_i^\varepsilon$, $i \geq 0$ is of order $1/\sqrt{\varepsilon}$.

Our next issue is the asymptotic behavior of the first eigenvalue $\lambda_0^\varepsilon$ as $\varepsilon \to 0$. We assume that the function $\varphi(s)$ in (4.1) still satisfies (4.2) and, in addition, $\xi(s)$ belongs to $\mathcal{C}^4([0,L])$. Recalling (2.23) and (3.4), we set:

$$\theta_0 := \frac{1}{2} M_0 \xi(s_0) \cdot \xi(s_0) + C_1\, \tilde{\tau}(s_0)^2 - C_2\, \tilde{\tau}'(s_0) + \frac{1}{16}\left(\frac{\rho_0 \cdot \xi^{(4)}(s_0)}{\rho_0 \cdot \xi''(s_0)}\right) - \frac{17}{9}\left(\frac{\rho_0 \cdot \xi^{(3)}(s_0)}{\rho_0 \cdot \xi''(s_0)}\right)^2. \quad (4.10)$$

**Conjecture.** *The first eigenvalue $\lambda_0^\varepsilon$ satisfies the following expansion*

$$\lambda_0^\varepsilon = \frac{\lambda_0}{\varepsilon^2} + \frac{\mu_0}{\varepsilon} + \frac{\nu_0}{\varepsilon^{1/2}} + \theta_0 + o(1).$$

In this paper we are able to prove the upper bound part of the conjecture above, namely

**Proposition 4.2.** *Let $\xi(s)$ belong to $\mathcal{C}^4([0,L])$ and satisfy (4.2). Then it holds*

$$\limsup_{\varepsilon \to 0}\left(\lambda_0^\varepsilon - \frac{\lambda_0}{\varepsilon^2} - \frac{\mu_0}{\varepsilon} - \frac{\nu_0}{\varepsilon^{1/2}}\right) \leq \theta_0. \quad (4.11)$$

We strongly believe that the upper bound obtained here is optimal. However, the proof of the lower bound inequality seems to require much more intricate arguments.

**Remark 4.3.** *In order to describe the localization of the eigenmodes $v_\varepsilon^i$ in the vicinity of $s_0$, we made a blow-up of function $v_\varepsilon^i$ by setting $\hat{v}_\varepsilon^i(t,y) := \varepsilon^{\alpha/2} v(s_0 + \varepsilon^\alpha t, y)$ where $\alpha = 1/4$ (notice that the $L^2$ norm remains unchanged). Let us explain this choice of $\alpha$ performing the change of variable $t := \dfrac{s - s_0}{\varepsilon^\alpha}$ in the shifted energy $J_\varepsilon(v) := \tilde{F}_\varepsilon(v) - (\lambda_0 + \varepsilon\,\mu_0)\int_Q \beta_\varepsilon |v|^2$, where in order to simplify we take $\tilde{\tau} = 0$ and $\gamma_0 = \gamma_1 = 0$ and in which we substitute the expression $\int_0^L \left[\int_\omega \beta_\varepsilon|\nabla_y v|^2 dy + \int_{\partial\omega}\beta_\varepsilon\,\gamma\,|v|^2 d\sigma\right] ds$, appearing in (2.14), by its optimal lower bound $\int_0^L \int_\omega \lambda(\varepsilon\xi)\beta_\varepsilon|v|^2 dy\, ds$. With $\hat{v}(t,y) := \varepsilon^{\alpha/2} v(s_0 + \varepsilon^\alpha t, y)$, $I_\varepsilon := \left[-\frac{s_0}{\varepsilon^\alpha}, \frac{L-s_0}{\varepsilon^\alpha}\right]$ and $\beta_\varepsilon \sim 1$, we roughly obtain*

$$J_\varepsilon(v) \sim \varepsilon^{2-2\alpha} \int_{I_\varepsilon}\int_\omega |\hat{v}'(t,y)|^2\, dy\, dt + \int_0^L \int_\omega \left(\lambda\bigl(\varepsilon\xi(s_0 + \varepsilon^\alpha t)\bigr) - (\lambda_0 + \varepsilon\,\mu_0)\right)|v|^2 dy\, ds.$$

*Exploiting (2.25) and (4.3), we can see that $\lambda\bigl(\varepsilon\xi(s_0 + \varepsilon^\alpha t)\bigr) - (\lambda_0 + \varepsilon\,\mu_0)$ is of order $\varepsilon^{1+2\alpha}$. Then, in order to balance the different powers of $\varepsilon$ in the expression above for $J_\varepsilon(v)$, we need that $1 + 2\alpha = 2 - 2\alpha$. Thus $\alpha = 1/4$ and $J_\varepsilon(v)$ is of order $\varepsilon^{3/4}$ which, after division by volume factor $\varepsilon^2$, gives exactly the exponent $\varepsilon^{-1/4}$ appearing in (4.9).*

In view of the discussion in Remark 4.3, we now fix the change of variables

$$t = \frac{s - s_0}{\varepsilon^{1/4}}, \quad \hat{v}(t) = \varepsilon^{1/8}\, v(s_0 + \varepsilon^{1/4}\, t), \quad t \in I_\varepsilon := \left[-\frac{s_0}{\varepsilon^{1/4}}, \frac{L - s_0}{\varepsilon^{1/4}}\right],$$

together with a rescaling of the energy, defining $\hat{G}_\varepsilon : L^2(\mathbb{R} \times \omega) \to \overline{\mathbb{R}}$ as follows

$$\hat{G}_\varepsilon(\hat{v}) := \begin{cases} \frac{1}{\varepsilon^{3/2}} \left( \tilde{F}_\varepsilon(v) - \int_{Q_L} \beta_\varepsilon (\lambda_0 + \varepsilon\mu_0)|v|^2 \right), & \text{if } v \in H^1(Q_L), \ v = 0 \text{ a.e. in } (\mathbb{R} \setminus [0,L]) \times \omega, \\ +\infty & \text{otherwise.} \end{cases}$$

Denoting $\bar{\beta}_\varepsilon(t) := \beta_\varepsilon(s_0 + \varepsilon^{1/4} t)$, $\bar{\tau}_\varepsilon(t) := \tilde{\tau}(s_0 + \varepsilon^{1/4} t)$, if $\hat{G}_\varepsilon(\hat{v}) < +\infty$, then

$$\begin{aligned}
\hat{G}_\varepsilon(\hat{v}) &= \int_{I_\varepsilon \times \omega} \frac{1}{\bar{\beta}_\varepsilon} |\hat{v}' + \varepsilon^{1/4}(\nabla_y \hat{v} \cdot R\, y)\bar{\tau}_\varepsilon|^2 \, dt\, dy \\
&\quad + \frac{1}{\varepsilon^{3/2}} \left[ \int_{I_\varepsilon \times \omega} \bar{\beta}_\varepsilon \left( |\nabla_y \hat{v}|^2 - (\lambda_0 + \varepsilon\mu_0)|\hat{v}|^2 \right) dt\, dy + \int_{I_\varepsilon \times \partial\omega} \bar{\beta}_\varepsilon \gamma |\hat{v}|^2 \, dt\, d\sigma \right] \\
&\quad + \varepsilon^{1/2} \int_{I_\varepsilon} \frac{\bar{\tau}_\varepsilon^2}{2} \left[ \int_{\partial\omega} \gamma\, (y \cdot \dot{y})^2 |\hat{v}|^2 \, d\sigma \right] dt \\
&\quad + \varepsilon^{1/2} \int_\omega \left( \gamma_0 \left| \hat{v}\left( -\frac{s_0}{\varepsilon^{1/4}}, y \right) \right|^2 + \gamma_L \left| \hat{v}\left( \frac{L-s_0}{\varepsilon^{1/4}}, y \right) \right|^2 \right) dy\,.
\end{aligned} \quad (4.12)$$

One checks that, for every $\varepsilon$, $v \in H^1(Q_L)$ if and only if $\hat{v} \in H^1(I_\varepsilon \times \omega)$ and it holds

$$\|v\|^2_{H^1(Q_L)} \leq C\, \varepsilon^{1/2}\, \|\hat{v}\|^2_{H^1(I_\varepsilon \times \omega)} \qquad (4.13)$$

In connection with the one dimensional spectral problem (4.5), we introduce the quadratic energy $\hat{G} : L^2(\mathbb{R} \times \omega) \to \overline{\mathbb{R}}$ defined as follows

$$\hat{G}(\hat{v}) := \begin{cases} \hat{G}_0(w) & \text{if } \hat{v}(t,y) = w(t)\, u_0(y), \ w \in H^1(\mathbb{R}) \cap L^2(\mathbb{R}; t^2 dt), \\ +\infty & \textit{otherwise}, \end{cases} \qquad (4.14)$$

where, with $\nu_0$ defined in (4.4),

$$\hat{G}_0(w) := \int_\mathbb{R} \left( |w'|^2 + t^2\, \nu_0^2\, |w|^2 \right) ds. \qquad (4.15)$$

As in the previous section the following proposition prepares our second main result

**Proposition 4.4.** *Under the assumptions of Theorem 4.1, the sequence of functionals $\{\hat{G}_\varepsilon\}$ defined in (4.12) satisfies all conditions i), ii) and iii) of Proposition 5.2, being the $\Gamma$-limit of $\hat{G}_\varepsilon$ given by (4.14) and (4.15).*

For subsequent estimates, it is useful to introduce

$$f_\varepsilon(t) := \frac{\Lambda_0\big(\varepsilon\, \xi(s_0 + \varepsilon^{1/4} t)\big) - (\lambda_0 + \varepsilon\, \mu_0)}{\varepsilon^{3/2}} \quad,\quad f_0(t) := \frac{t^2}{2}\left(\rho_0 \cdot \xi''(s_0)\right)\ (= \nu_0^2\, t^2) \qquad (4.16)$$

**Lemma 4.5.** *Let $\xi(s)$ be of class $\mathcal{C}^2([0,L])$ and satisfy (4.2) and (4.3). Then, for small $\varepsilon$, there holds*

$$f_\varepsilon(t) \geq \eta_0\, t^2 - c\sqrt{\varepsilon} \qquad (4.17)$$

*Moreover, the convergence $f_\varepsilon \to f_0$ holds uniformly on bounded subsets of $\mathbb{R}$.*

*Proof.* By (2.25) and the boundedness of function $\xi(s)$, it holds $\left|\Lambda_0\bigl(\varepsilon\xi(s)\bigr) - (\lambda_0 + \varepsilon\,\varphi(s))\right| \leq C\,\varepsilon^2$. Since by (4.3) we have $\varphi(s_0 + \varepsilon^{1/4}t) \geq \mu_0 + \varepsilon^{1/2}\eta_0 t^2$, it follows that

$$\Lambda_0\bigl(\varepsilon\xi(s_0+\varepsilon^{1/4}t)\bigr) - (\lambda_0+\varepsilon\,\mu_0) \geq \varepsilon^{3/2}\eta_0\, t^2 - C\,\varepsilon^2\,,$$

which, after dividing by $\varepsilon^{3/2}$, leads to lower bound (4.17). Moreover, for every $t$, we write the following Taylor expansion (at third order in $\varepsilon^{1/4}$): $\xi(s_0+\varepsilon^{1/4}t) = \xi(s_0) + \varepsilon^{1/4}t\,\xi'(s_0) + \varepsilon^{1/2}\frac{t^2}{2}\,\xi''(s_0+\varepsilon^{1/4}\theta\,t)$, being $\theta$ a suitable value in $(0,1)$. Then, inserting in (2.25), we are led to

$$|f_\varepsilon(t) - f_0(t)| \leq \frac{t^2}{2}\left|\bigl(\varphi''(s_0+\varepsilon^{1/4}\theta\,t)\bigr) - \varphi''(s_0)\right| + C\,\varepsilon^{3/2}\,,$$

and $f_\varepsilon \to f_0$ uniformly on compact subsets, thanks to the uniform continuity of $\varphi''$ on $[0,L]$. $\square$

**Proof of Proposition 4.4.** The conditions $i)$, $ii)$ of Proposition (5.2) are established in Step 1. The $\Gamma-$ convergence of $\hat G_\varepsilon$ (condition $iii)$) is proved by checking the lower bound inequality in Step 2 and the upper bound inequality in Step 3.

**Step 1.** *(Coercivity and compactness)* Looking at the expression (4.12), since $\gamma, \gamma_0, \gamma_L$ are non negative (see (2.11)), we obtain

$$\begin{aligned}\hat G_\varepsilon(\hat v) &\geq \int_\omega\!\!\int_{I_\varepsilon}\frac{1}{\bar\beta_\varepsilon}|\hat v' + \varepsilon^{1/4}(\nabla_y\hat v\cdot R\ y)\bar\tau_\varepsilon|^2\,dt\,dy \\ &+ \frac{1}{\varepsilon^{3/2}}\left[\int_\omega\!\!\int_{I_\varepsilon}\bar\beta_\varepsilon\Bigl(|\nabla_y\hat v|^2 - (\lambda_0+\varepsilon\mu_0)|\hat v|^2\Bigr)\,dt\,dy + \int_{\partial\omega}\!\!\int_{I_\varepsilon}\bar\beta_\varepsilon\gamma|\hat v|^2\,dt\,d\sigma\right].\end{aligned}$$
(4.18)

On the other hand, by (2.16), (2.19) and (4.16), we have the sharp lower bound

$$\frac{1}{\varepsilon^{3/2}}\left[\int_\omega\!\!\int_{I_\varepsilon}\bar\beta_\varepsilon\Bigl(|\nabla_y\hat v|^2 - (\lambda_0+\varepsilon\mu_0)|\hat v|^2\Bigr)\,dt\,dy + \int_{\partial\omega}\!\!\int_{I_\varepsilon}\bar\beta_\varepsilon\gamma|\hat v|^2\,dt\,d\sigma\right] \geq \int_\omega\!\!\int_{I_\varepsilon} f_\varepsilon\,\bar\beta_\varepsilon\,|\hat v|^2\,dt\,dy,$$

from which follows that

$$\hat G_\varepsilon(\hat v) \geq \int_\omega\!\!\int_{I_\varepsilon}\left[\frac{1}{\bar\beta_\varepsilon}|\hat v' + \varepsilon^{1/4}(\nabla_y\hat v\cdot R\ y)\bar\tau_\varepsilon|^2 + \bar\beta_\varepsilon\,f_\varepsilon\,|\hat v|^2\right]\,dt\,dy\,. \quad (4.19)$$

In particular, for small $\varepsilon$, as $\bar\beta_\varepsilon \geq 1/2$ whereas $f_\varepsilon \geq -c$ by (4.17), the condition $i)$ of Proposition 5.2 is fulfilled by $\hat G_\varepsilon$.

In order to check condition $ii)$, consider a bounded sequence $\{\hat v_\varepsilon\}$ in $L^2(\mathbb{R}\times\omega)$ satisfying $\sup \hat G_\varepsilon(\hat v_\varepsilon) < +\infty$. Then, in view of (4.17), (4.18) and (4.19), we obtain

$$\int_{I_\varepsilon\times\omega}|\nabla_y\hat v_\varepsilon|^2 \leq M\,,\quad \int_{I_\varepsilon\times\omega}\left|\hat v'_\varepsilon + \varepsilon^{1/4}(\nabla_y\hat v_\varepsilon\cdot R\ y)\,\bar\tau_\varepsilon\right|^2 \leq M\,,\quad \int_{I_\varepsilon\times\omega}t^2|\hat v_\varepsilon|^2\,dt\,dy \leq M,\ (4.20)$$

for a suitable constant $M$ independent of $\varepsilon$ (notice that, by definition, the condition $\hat G_\varepsilon(\hat v_\varepsilon) < +\infty$ implies $\hat v_\varepsilon = 0$ a.e. in $(\mathbb{R}\setminus I_\varepsilon)\times\omega$). Possibly after extracting a subsequence of $\{\hat v_\varepsilon\}$, we may assume that $\hat v_\varepsilon \rightharpoonup \hat v$ weakly in $L^2(\mathbb{R}\times\omega)$. Then, by (4.20), for every finite $\eta > 0$, $\{\hat v_\varepsilon\}$ is weakly compact in $H^1((-\eta,\eta)\times\omega))$ so that by Rellich-Kondrachov Theorem, there holds

$$\lim_\varepsilon \int_{(-\eta,\eta)\times\omega}|\hat v_\varepsilon|^2 = \int_{(-\eta,\eta)\times\omega}|\hat v|^2\,.$$

By exploiting the third inequality in (4.20) and as $\hat v_\varepsilon = 0$ outside $I_\varepsilon$, we obtain

$$\int_{\mathbb{R}\times\omega} |\hat{v}|^2 \, dt \, dy \;\leq\; \liminf_{\varepsilon\to 0} \int_{I_\varepsilon \times\omega} |\hat{v}_\varepsilon|^2 \, dt \, dy \;\leq\; \limsup_{\varepsilon\to 0} \int_{I_\varepsilon \times\omega} |\hat{v}_\varepsilon|^2 \, dt \, dy \;\leq\; \int_{|t|\leq \eta} |\hat{v}|^2 \, dt \, dy \;+\; \frac{M}{\eta^2} \;,$$

from which folllows the strong convergence of $\hat{v}_\varepsilon$ in $L^2(\mathbb{R}\times\omega)$ by sending $\eta$ to infinity.

**Step 2.** *(Lower bound inequality).* Let $\{\hat{v}_\varepsilon\}$ be a sequence such that $\hat{v}_\varepsilon \to \hat{v}$ in $L^2(\mathbb{R}\times\omega)$. We have to establish

$$\liminf \hat{G}_\varepsilon(\hat{v}_\varepsilon) \;\geq\; \hat{G}(\hat{v}) \;. \qquad (4.21)$$

Up to a subsequence we may assume that $\liminf_{\varepsilon\to 0} \hat{G}_\varepsilon(\hat{v}_\varepsilon) = \lim_{\varepsilon\to 0} \hat{G}_\varepsilon(\hat{v}_\varepsilon) < +\infty$. Then, as noticed in Step 1, the sequence $\{\hat{v}_\varepsilon\}$ is bounded in $H^1_{\text{loc}}(\mathbb{R}\times\omega)$ since estimates (4.20) hold. Therefore, the limit $\hat{v}$ is an element of $H^1_{\text{loc}}(\mathbb{R}\times\omega)$. Let us pass to the lower limit in inequality (4.19): since $\bar\beta_\varepsilon \to 1$ uniformly, while $\bar\tau_\varepsilon$ remains bounded, and $f_\varepsilon$ converges pointwise to $f_0$ and satisfies a uniform lower bound (see Lemma 4.5), with the help of Fatou's Lemma we obtain

$$\liminf_{\varepsilon\to 0} \hat{G}_\varepsilon(\hat{v}_\varepsilon) \;\geq\; \int_{\mathbb{R}\times\omega} \left(|\hat{v}'|^2 + f_0(t)|\hat{v}|^2 dt\right) \, dy \;. \qquad (4.22)$$

On the other hand, from (4.18) and since $\hat{G}_\varepsilon(\hat{v}_\varepsilon)$ is uniformly bounded, one has

$$\int_{I_\varepsilon \times\omega} \bar\beta_\varepsilon \left(|\nabla_y \hat{v}_\varepsilon|^2 - (\lambda_0 + \varepsilon\mu_0)|\hat{v}_\varepsilon|^2\right) \, dt \, dy \;+\; \int_{I_\varepsilon \times\partial\omega} \bar\beta_\varepsilon \, \gamma(y)\,|\hat{v}_\varepsilon|^2 \, dt\, d\sigma(y) \;\leq\; C\,\varepsilon^{3/2} \;.$$

It follows that the function $h_\varepsilon(t) := \int_\omega \left(|\nabla_y \hat{v}_\varepsilon(t,\cdot)|^2 - \lambda_0|\hat{v}_\varepsilon(t,\cdot)|^2\right) \, dy + \int_{\partial\omega} \gamma(y)\,|\hat{v}_\varepsilon(t,\cdot)|^2 \, d\sigma(y)$, which by the definition of $\lambda_0$ is nonnegative, does converge to zero in $L^1_{\text{loc}}(\mathbb{R})$. We notice that the function $h_0(t) := \int_\omega \left(|\nabla_y \hat{v}(t,\cdot)|^2 - \lambda_0|\hat{v}(t,\cdot)|^2\right) \, dy + \int_{\partial\omega} \gamma(y)\,|\hat{v}(t,\cdot)|^2 \, d\sigma(y)$ is nonnegative as well. Moreover, since $\hat{v}_\varepsilon \to \hat{v}$ strongly in $L^2(\mathbb{R}\times\omega)$ and weakly in $H^1_{\text{loc}}(\mathbb{R}\times\omega)$, for every $R>0$, there holds $0 = \liminf_{\varepsilon\to 0} \int_{|t|<R} h_\varepsilon(t)\, dt \;\geq\; \int_{|t|<R} h_0(t)\, dt$. This implies that $h_0(t) = 0$ a.e. and therefore $\hat{v}(t,\cdot)$ is an eigenvector associated with $\lambda_0$ (see (2.20)). It follows that $\hat{v}$ can be written in the form $\hat{v}(t,y) = w(t)\,u_0(y)$ with $w \in H^1_{\text{loc}}(\mathbb{R})$. Plugging this expression of $\hat{v}$ into the right hand side of (4.22) we obtain that $w$ belongs to $H^1(\mathbb{R}) \cap L^2(\mathbb{R}; t^2 dt)$ and, in view of (4.14) and (4.15), we see that (4.22) is nothing else but the lower bound inequality (4.21).

**Step 3** *(Upper bound inequality).* Let $\hat{v} \in L^2(\mathbb{R}\times\omega)$. We have to construct a sequence $\{\hat{v}_\varepsilon\}$ such that $\hat{v}_\varepsilon \to \hat{v}$ and $\limsup_{\varepsilon\to 0} \hat{G}_\varepsilon(\hat{v}_\varepsilon) \leq \hat{G}(\hat{v})$. We may assume that $\hat{G}(\hat{v}) < +\infty$ so that we can write $\hat{v}(t,y) = \hat{w}(t)\,u_0(y)$ for a suitable element $\hat{w} \in H^1(\mathbb{R}) \cap L^2(\mathbb{R}; t^2 dt)$.

We consider $\hat{v}_\varepsilon$ defined by $\hat{v}_\varepsilon = \chi_\varepsilon(t)\,\hat{w}(t)\,u_0(y)$, being $\chi_\varepsilon$ on $I_\varepsilon \times\omega$. Then, substituting in the formula (4.12) and taking into account that $w$ is bounded while $\bar\beta_\varepsilon \to 1$ uniformly and $\int_\omega u_0^2 = 1$, we infer, after some computations, that

$$\limsup_{\varepsilon\to 0} \hat{G}_\varepsilon(\hat{v}_\varepsilon) \;=\; \int_\mathbb{R} |\hat{w}'|^2 \, dt \;+\; \limsup_{\varepsilon\to 0} \frac{1}{\varepsilon^{3/2}} \int_{I_\varepsilon} \Phi_\varepsilon\!\left(\xi(s_0 + t\,\varepsilon^{1/4})\right) |\hat{w}|^2 \, dt \;,$$

where for every $\xi \in \mathbb{R}^2$ we have set:

$$\Phi_\varepsilon(\xi) \;:=\; \int_\omega (1-\varepsilon\xi\cdot y)\Big(|\nabla_y u_0|^2 - (\lambda_0+\varepsilon\mu_0)|u_0|^2\Big) \, dy \;+\; \int_{\partial\omega} (1-\varepsilon\xi\cdot y)\,\gamma\,|u_0|^2 \, d\sigma \;.$$

In view of (4.15), we are reduced to show that

$$\limsup_{\varepsilon\to 0} \frac{1}{\varepsilon^{3/2}} \int_{I_\varepsilon} \Phi_\varepsilon\!\left(\xi(s_0 + t\,\varepsilon^{1/4})\right) |\hat{w}|^2 \, dt \;\leq\; \int_\mathbb{R} \nu_0^2 \, t^2 \, |\hat{w}|^2 \, dt \;. \qquad (4.23)$$

Since $u_0$ satisfy (2.20), the $\varepsilon^0$ order term in $\Phi_\varepsilon(\xi)$ vanishes. By writing relation (2.36) with $\psi = u_0$ and recalling (2.21), we get $\Phi_\varepsilon(\xi) = \varepsilon\, (\xi \cdot \rho_0 - \mu_0) + \varepsilon^2 \mu_0 \int_\omega \xi \cdot y \, dy$. Thus, by (4.2), we have the estimate

$$\left| \frac{1}{\varepsilon^{3/2}} \Phi_\varepsilon\left(\xi(s_0 + t\,\varepsilon^{1/4})\right) - \varphi_\varepsilon(t) \right| \leq C\sqrt{\varepsilon} \quad \text{where} \quad \varphi_\varepsilon(t) := \frac{\varphi((s_0 + t\,\varepsilon^{1/4}) - \varphi(s_0)}{\varepsilon^{1/2}}.$$

Therefore, the concluding inequality (4.23) is achieved provided

$$\lim_{\varepsilon \to 0} \int_{I_\varepsilon} \varphi_\varepsilon(t) \, |\hat{w}|^2 \, dt = \int_{\mathbb{R}} (\nu_0)^2 \, t^2 \, |\hat{w}|^2 \, dt \, .$$

This is a consequence of the dominated convergence Theorem, since $\varphi_\varepsilon(t) \to (\nu_0)^2 t^2$ and, by (4.3), it holds $|\varphi_\varepsilon(t)| \leq \frac{1}{\eta_0} t^2$ whereas $\int_{\mathbb{R}} t^2 |\hat{w}|^2 \, dt < +\infty$. □

**Proof of Theorem 4.1** We observe that, by (2.15) and (4.13), there holds for every $u \in H^1(\Omega_\varepsilon)$

$$\left| \frac{1}{\varepsilon^{3/2}} \left[ F_\varepsilon(u) - (\lambda_0 + \varepsilon\mu_0) \int_{\Omega_\varepsilon} u^2 \, dx \right] - \hat{G}_\varepsilon(\hat{v}) \right| \leq C\, \varepsilon^{3/2} \, \|v\|^2_{H^1(Q_L)} \leq C'\, \varepsilon \, \|\hat{v}\|^2_{H^1(I_\varepsilon \times \omega)} \quad (4.24)$$

where $v(s,y) = u \circ \psi_\varepsilon(s,y)$ and $\hat{v}(t,y) = \varepsilon^{1/8} v(s_0 + \varepsilon^{1/4} t)$. The asymptotic behavior of $\nu_\varepsilon^i = \frac{1}{\varepsilon^{3/2}} [\lambda_i^\varepsilon - (\lambda_0 + \varepsilon\mu_0)]$ is therefore ruled by the functional $\hat{G}_\varepsilon$ to which we apply Proposition 4.4. The proof follows by using exactly the same line as in the proof of Theorem 3.1. □

**Proof of Proposition 4.2** We introduce $\hat{H}_\varepsilon : L^2(\mathbb{R} \times \omega) \to \overline{\mathbb{R}}$ defined by

$$\hat{H}_\varepsilon(\hat{v}) := \frac{1}{\varepsilon^{1/2}} \hat{G}_\varepsilon(\hat{v}) - \frac{1}{\varepsilon^{1/2}} \int_{I_\varepsilon \times \omega} \bar{\beta}_\varepsilon \, \nu_0 |\hat{v}|^2 dt dy \, .$$

For $\hat{v} \in H^1(I_\varepsilon \times \omega)$ and $\hat{v}$ vanishing in $(\mathbb{R} \setminus I_\varepsilon) \times \omega$, the expression of $\hat{H}_\varepsilon(\hat{v})$ reads

$$\hat{H}_\varepsilon(\hat{v}) = \hat{A}_\varepsilon(\hat{v}) + \hat{B}_\varepsilon(\hat{v}) + \hat{C}_\varepsilon(\hat{v}), \quad \text{where}$$

$$\hat{A}_\varepsilon(\hat{v}) := \int_{I_\varepsilon \times \omega} \frac{1}{\bar{\beta}_\varepsilon} \left| \frac{1}{\varepsilon^{1/4}} \hat{v}' + (\nabla_y \hat{v} \cdot R\, y) \bar{\tau}_\varepsilon \right|^2 dt \, dy$$

$$\hat{B}_\varepsilon(\hat{v}) := \frac{1}{\varepsilon^2} \left[ \int_{I_\varepsilon \times \omega} \bar{\beta}_\varepsilon \left( |\nabla_y \hat{v}|^2 - (\lambda_0 + \varepsilon\mu_0 + \varepsilon^{3/2}\nu_0)|\hat{v}|^2 \right) dt \, dy + \int_{I_\varepsilon \times \partial\omega} \bar{\beta}_\varepsilon \gamma |\hat{v}|^2 \, dt \, d\sigma \right]$$

$$\hat{C}_\varepsilon(\hat{v}) := \int_{I_\varepsilon \times \partial\omega} \frac{\bar{\tau}_\varepsilon^2}{2} \gamma\, (y \cdot \dot{y})^2 |\hat{v}|^2 \, dt \, d\sigma + \int_\omega \left( \gamma_0 \left| \hat{v}\left(\frac{-s_0}{\varepsilon^{1/4}}, y\right) \right|^2 + \gamma_L \left| \hat{v}\left(\frac{L - s_0}{\varepsilon^{1/4}}, y\right) \right|^2 \right) dy$$

(4.25)

Setting

$$\theta^\varepsilon := \inf \left\{ \frac{\hat{H}_\varepsilon(\hat{v})}{\int_{\mathbb{R} \times \omega} \bar{\beta}_\varepsilon |\hat{v}|^2 \, dt dy} \right\}, \quad (4.26)$$

we claim that

$$|\theta^\varepsilon - \theta_0^\varepsilon| \leq C \sqrt{\varepsilon} \quad \text{where} \quad \theta_0^\varepsilon := \lambda_0^\varepsilon - \frac{\lambda_0}{\varepsilon^2} - \frac{\mu_0}{\varepsilon} - \frac{\nu_0}{\varepsilon^{1/2}}. \quad (4.27)$$

Indeed, by dividing the inequality (4.24) by $\sqrt{\varepsilon}$, we infer that for every $u \in H^1(\Omega_\varepsilon)$ it holds

$$\left| \frac{1}{\varepsilon^2} \left[ F_\varepsilon(u) - (\lambda_0 + \varepsilon\mu_0 + \varepsilon^{3/2}\nu_0) \int_{\Omega_\varepsilon} |u|^2 \, dx \right] - \hat{H}_\varepsilon(\hat{v}) \right| \leq \sqrt{\varepsilon} \, \|\hat{v}\|^2_{H^1(I_\varepsilon \times \Omega)},$$

being $\hat{v}(t,y) = \varepsilon^{1/8} u \circ \psi_\varepsilon(s_0 + \varepsilon^{1/4}t, y)$ in $I_\varepsilon \times \omega$ and zero in $(\mathbb{R} \setminus I_\varepsilon) \times \omega$. The claim follows by comparing the Rayleigh quotients associated with $\theta^\varepsilon$ and $\theta_0^\varepsilon$, respectively.

Let $\bar{\xi}_\varepsilon(t) := \xi(s_0 + \varepsilon^{1/4}t)$ and let $\chi_{\bar{\xi}_\varepsilon(t)}$ be the solution of (2.22) for $z = \bar{\xi}_\varepsilon(t)$. We consider the approximating sequence $(\hat{v}_\varepsilon)$ defined on $I_\varepsilon \times \omega$ as follows

$$\hat{v}_\varepsilon(t,y) := \left(\hat{w}_0(t) + \varepsilon^{1/4}\hat{\varphi}(t)\right)\left(u_0(y) + \varepsilon \chi_{\bar{\xi}_\varepsilon(t)}(y)\right), \quad (4.28)$$

and zero outside $I_\varepsilon \times \omega$, where the function $\hat{\varphi}$, specified later, will be a suitable linear combination of the eigenfunctions $\{\hat{w}_i, i = 1, 3\}$ defined in (4.7). In particular, in order that the total energy $\hat{H}_\varepsilon(\hat{v}_\varepsilon)$ remains finite, we will need that $\int_\mathbb{R} \hat{\varphi}\hat{w}_0 = \int_\mathbb{R} t^2\hat{\varphi}\hat{w}_0 = 0$. Thanks to the normalization condition on functions $u_0, \hat{w}_0$, and recalling that $\bar{\beta}_\varepsilon \to 1$ uniformly, one checks that $\lim\limits_{\varepsilon \to 0} \int_{\mathbb{R} \times \omega} \bar{\beta}_\varepsilon |\hat{v}_\varepsilon|^2 \, dt \, dy = 1$. Thus, by (4.26) and (4.27), the upper bound inequality of Proposition (4.2) is established once we have shown that

$$\limsup_{\varepsilon \to 0} \hat{H}_\varepsilon(\hat{v}_\varepsilon) \leq \theta_0. \quad (4.29)$$

We will establish successively the following convergences:

$$\lim_{\varepsilon \to 0}\left(\hat{A}_\varepsilon(\hat{v}_\varepsilon) - \frac{\nu_0}{2\sqrt{\varepsilon}}\int_{I_\varepsilon}|\hat{w}_0|^2 - \frac{2}{\varepsilon^{1/4}}\int_{I_\varepsilon}\hat{w}_0'\hat{\varphi}'\right) = \tilde{\tau}^2(s_0)\int_\omega |\nabla u_0 \cdot Ry|^2 - C_2\,\tilde{\tau}'(s_0) + \int_\mathbb{R}|\hat{\varphi}'|^2\,dt \quad (4.30)$$

$$\lim_{\varepsilon \to 0}\left(\hat{B}_\varepsilon(\hat{v}_\varepsilon) + \frac{\nu_0}{2\sqrt{\varepsilon}}\int_{I_\varepsilon}|\hat{w}_0|^2 + \frac{2}{\varepsilon^{1/4}}\int_{I_\varepsilon}\hat{w}_0'\hat{\varphi}'\right) = \frac{1}{2}M_0\,\xi(s_0)\cdot\xi(s_0) + \frac{\rho_0\cdot\xi^{(4)}(s_0)}{32(\nu_0)^2} +$$
$$+ \int_\mathbb{R}(\nu_0^2 t^2 - \nu_0)|\hat{\varphi}|^2\,dt + \int_\mathbb{R}\frac{\rho_0\cdot\xi^{(3)}(s_0)}{3}t^3\,\hat{w}_0\,\hat{\varphi}\,dt \quad (4.31)$$

$$\lim_{\varepsilon \to 0}\hat{C}_\varepsilon(\hat{v}_\varepsilon) = \frac{1}{2}\tilde{\tau}^2(s_0)\int_{\partial\omega}\gamma|u_0|^2(y\cdot\dot{y})^2\,d\sigma. \quad (4.32)$$

Adding up the three previous equalities, we infer that

$$\limsup_{\varepsilon \to 0}\hat{H}_\varepsilon(\hat{v}_\varepsilon) \leq \frac{1}{2}M_0\xi(s_0)\cdot\xi(s_0) + C_1\,\tilde{\tau}^2(s_0) - C_2\,\tilde{\tau}'(s_0) + \frac{\rho_0\cdot\xi^{(4)}(s_0)}{32\,\nu_0^2} + \mathcal{E}(\hat{\varphi}) \quad (4.33)$$

where the last term, to be minimized with respect to $\hat{\varphi} \in H^1(\mathbb{R})$, is given by

$$\mathcal{E}(\hat{\varphi}) := \int_\mathbb{R}|\hat{\varphi}'|^2\,dt + \int_\mathbb{R}\left[\nu_0^2 t^2 - \nu_0\right]|\hat{\varphi}|^2\,dt + 2\int_\mathbb{R}h(t)\,\hat{\varphi}\,dt \quad \text{with} \quad h(t) := \frac{\rho_0\cdot\xi^{(3)}(s_0)}{6}t^3\,\hat{w}_0.$$

It turns out that $h$ is orthogonal to $\hat{w}_0$ and can be expressed as a linear combination of normalized eigenvectors $\hat{w}_1, \hat{w}_3$ introduced in (4.7). In fact, we have

$$\frac{\hat{w}_1}{\hat{w}_0}(t) = (\nu_0)^{1/2}\,\frac{t}{\sqrt{2}} \quad , \quad \frac{\hat{w}_3}{\hat{w}_0}(t) = (\nu_0)^{3/2}\,\frac{t^3}{4\sqrt{3}} - (\nu_0)^{1/2}\,\frac{\sqrt{3}}{4}\,t,$$

and, consequently,

$$h(t) = \frac{\rho_0\cdot\xi^{(3)}(s_0)}{(\nu_0)^{3/2}}\left(\frac{1}{\sqrt{2}}\,\hat{w}_1(t) + \frac{2}{\sqrt{3}}\,\hat{w}_3(t)\right).$$

Applying Lemma 5.1, we deduce that the minimum of $\mathcal{E}(\hat{\varphi})$ is reached for a suitable linear combination $\hat{\varphi}_{\text{opt}}$ of $\hat{w}_1$ and $\hat{w}_3$. Taking into account that $\nu_3 - \nu_0 = 6\nu_0$, $\nu_1 - \nu_0 = 2\nu_0$ and recalling (4.4), we have

$$\min \mathcal{E} = -\frac{17}{36}\left[\frac{\rho_0 \cdot \xi^{(3)}(s_0)}{(\nu_0)^4}\right]^2 = -\frac{17}{9}\left[\frac{\rho_0 \cdot \xi^{(3)}(s_0)}{\rho_0 \cdot \xi''(s_0)}\right]^2.$$

Then, plugging $\hat{\varphi} = \hat{\varphi}_{\text{opt}}$ into the definition (4.28) of $\hat{v}_\varepsilon$ and in view of (4.10), the upper bound inequality (4.29) follows directly from (4.33).

It remains to show claims (4.30), (4.31) and (4.32).

Let us plug $\hat{v}_\varepsilon$ into the three expressions in (4.25), taking into account that for all $t \in I_\varepsilon$ it holds

$$\hat{v}'_\varepsilon(t, y) = (\hat{w}'_0 + \varepsilon^{1/4}\hat{\varphi}')(t)\,(u_0 + \varepsilon\,\chi_{\bar{\xi}_\varepsilon})(y) + \varepsilon^{5/4}\,\hat{w}_0(t)\,\chi_{\bar{\xi}_\varepsilon}\,(\bar{\xi}_\varepsilon)'(t)\,,\quad \nabla_y \hat{v}_\varepsilon(t) = (\hat{w}_0 + \varepsilon^{1/4}\hat{\varphi})(t)\,(\nabla_y u_0 + \varepsilon\,\nabla_y \chi_{\bar{\xi}_\varepsilon}).$$

In particular, we have the following estimates:

$$\|\hat{v}'_\varepsilon(t,y) - (\hat{w}'_0 + \varepsilon^{1/4}\hat{\varphi}')(t)\,u_0(y)\|_{L^2(I_\varepsilon \times \omega)} \leq C\,\varepsilon\quad,\quad \|\nabla_y \hat{v}_\varepsilon(t,y) - (\hat{w}_0 + \varepsilon^{1/4}\hat{\varphi})(t)\,\nabla_y u_0(y)\|_{L^2(I_\varepsilon \times \omega)} \leq C\,\varepsilon\,.$$

Thus, as $|1 - \beta_\varepsilon| \leq C\varepsilon$ and observing that, by (4.8) and the exponential decay of $\hat{w}_0$, we have $\left|\int_{I_\varepsilon}(|\hat{w}'_0|^2 - \frac{\nu_0}{2}|\hat{w}_0|^2)\right| \ll \sqrt{\varepsilon}$, it follows that the left hand side of (4.30) has the same asymptotic behavior as

$$L_\varepsilon := \frac{1}{\sqrt{\varepsilon}}\left[\int_{I_\varepsilon \times \omega}\left|(\hat{w}'_0 + \varepsilon^{1/4}\hat{\varphi}')(t)u_0(y) + \varepsilon^{1/4}\,(\hat{w}_0 + \varepsilon^{1/4}\hat{\varphi})(t)\,(\nabla_y u_0 \cdot R\,y))\bar{\tau}_\varepsilon\right|^2 - \int_{I_\varepsilon}(|\hat{w}'_0|^2 + 2\varepsilon^{1/4}\hat{w}'_0 \hat{\varphi}')\right].$$

After straightforward computations, taking into account that $\bar{\tau}_\varepsilon \to \tilde{\tau}(s_0)$ uniformly and integrating with respect to $y \in \omega$, we obtain

$$\lim_{\varepsilon \to 0} L_\varepsilon = \tilde{\tau}^2(s_0)\int_\omega |\nabla u_0 \cdot Ry|^2 + \lim_{\varepsilon \to 0}\left[J_\varepsilon(\hat{\varphi}) + \frac{r_\varepsilon}{\varepsilon^{1/4}}\right]\,, \tag{4.34}$$

where $J_\varepsilon(\hat{\varphi}) := \int_{I_\varepsilon}(|\hat{\varphi}'|^2 + 2\,C_2\int_{I_\varepsilon}(\hat{w}'_0\hat{\varphi} + \hat{w}_0\hat{\varphi}')\,\bar{\tau}_\varepsilon\,dt$ and $r_\varepsilon := 2\,C_2 \int_{I_\varepsilon}(\hat{w}'_0(t)\hat{w}_0(t))\,\bar{\tau}_\varepsilon\,dt$.

Integrating by parts, and as the boundary terms are exponentially small ($\hat{\varphi}$ is a linear combination of $\hat{w}_1, \hat{w}_3$), it holds that

$$\lim_{\varepsilon \to 0}\int_{I_\varepsilon}(\hat{w}'_0\hat{\varphi} + \hat{w}_0\hat{\varphi}')\,\bar{\tau}_\varepsilon\,dt = -\lim_{\varepsilon \to 0}\int_{I_\varepsilon}\hat{w}_0\hat{\varphi}\,\tilde{\tau}'(s_0 + \varepsilon^{1/4}t)\,\varepsilon^{1/4}\,dt = 0\,,$$

$$\lim_{\varepsilon \to 0}\int_{I_\varepsilon}|\hat{w}'_0|^2\,\bar{\tau}_\varepsilon\,dt = -\lim_{\varepsilon \to 0}\int_{I_\varepsilon}|\hat{w}_0|^2\,\tilde{\tau}'(s_0 + \varepsilon^{1/4}t)\,\varepsilon^{1/4}\,dt = -\tilde{\tau}'(s_0)\,,$$

where in the last integral we use dominated convergence. Therefore, we obtain that

$$\lim_{\varepsilon \to 0} r_\varepsilon\,\varepsilon^{-1/4} = -C_2\,\tilde{\tau}'(s_0)\quad,\quad \lim_{\varepsilon \to 0} J_\varepsilon(\hat{\varphi}) = \int_{\mathbb{R}}|\hat{\varphi}'|^2\,dt\,.$$

Then (4.30) follows from (4.34).

The derivation of (4.32) is straightforward since

$$\hat{C}_\varepsilon(\hat{v}_\varepsilon) = \frac{1}{2}\left(\int_{\partial\omega}\gamma|u_0|^2(y \cdot \dot{y})^2\,d\sigma\right)\left(\int_{I_\varepsilon}\bar{\tau}_\varepsilon^2\,|\hat{w}_0|^2\,dt\right)$$

$$+ \left(\int_\omega \gamma_0|u_0|^2\,dy\right)\left|\hat{w}_0\left(-\frac{s_0}{\varepsilon^{1/4}}\right)\right|^2 + \left(\int_\omega \gamma_L|u_0|^2\,dy\right)\left|\hat{w}_0\left(\frac{L - s_0}{\varepsilon^{1/4}}\right)\right|^2,$$

where the expressions in the last line vanish as $\varepsilon \to 0$ due the exponential decay of $\hat{w}_0$.

Eventually we finish the proof by establishing the claim 4.31. Let us insert $\beta_\varepsilon = 1 - \varepsilon(\bar{\xi}_\varepsilon(t)\cdot y)$ and $\hat{v}_\varepsilon$ given by (4.28) in the expression of $\hat{B}_\varepsilon$ (see (4.25)). By using the assertion ii) of Proposition 2.1, we have

$$\lim_{\varepsilon\to 0} \frac{1}{\varepsilon^2} \left[ \int_{I_\varepsilon\times\omega} \bar{\beta}_\varepsilon \left( |\nabla_y \hat{v}_\varepsilon|^2 - (\lambda_0 + \varepsilon\rho_0\cdot\bar{\xi}_\varepsilon(t))|\hat{v}_\varepsilon|^2 \right) dt\,dy + \int_{I_\varepsilon\times\partial\omega} \bar{\beta}_\varepsilon \gamma |\hat{v}_\varepsilon|^2 \, dt\,d\sigma \right]$$

$$= \lim_{\varepsilon\to 0} \frac{1}{\varepsilon^2} \int_{I_\varepsilon} |\hat{w}_0 + \varepsilon^{1/4}\hat{\varphi}|^2 \, E_{\varepsilon\bar{\xi}_\varepsilon(t)}(u_0 + \chi_{\varepsilon\bar{\xi}_\varepsilon(t)}) \, dt$$

$$= \lim_{\varepsilon\to 0} \int_{I_\varepsilon} \frac{1}{2} \left( M_0\,\bar{\xi}_\varepsilon(t) \cdot \bar{\xi}_\varepsilon(t) \right) |\hat{w}_0 + \varepsilon^{1/4}\hat{\varphi}|^2 \, dt$$

$$= \frac{1}{2} M_0\, \xi(s_0) \cdot \xi(s_0)$$

where in the last line we used dominated convergence, the exponential decay of $\hat{w}_0$ and $\int_\mathbb{R} |\hat{w}_0|^2 = 1$. Thus, recalling the definition of $\hat{B}_\varepsilon$ in (4.25), we deduce that

$$\lim_{\varepsilon\to 0} \left( \hat{B}_\varepsilon(\hat{v}_\varepsilon) + \frac{\nu_0}{2\sqrt{\varepsilon}}\int_{I_\varepsilon} |\hat{w}_0|^2 + \frac{2}{\varepsilon^{1/4}}\int_{I_\varepsilon} \hat{w}_0'\hat{\varphi}' \right) = \frac{1}{2} M_0\,\xi(s_0)\cdot\xi(s_0) + \lim_{\varepsilon\to 0} U_\varepsilon(\hat{\varphi}) \,, \qquad (4.35)$$

where we have set

$$U_\varepsilon(\hat{\varphi}) := \int_{I_\varepsilon} \left( \frac{(\rho_0\cdot\bar{\xi}_\varepsilon) - \mu_0 - \nu_0\sqrt{\varepsilon}}{\varepsilon} \right) |\hat{w}_0 + \varepsilon^{1/4}\hat{\varphi}|^2 \, dt + \frac{\nu_0}{2\sqrt{\varepsilon}} \int_{I_\varepsilon} |\hat{w}_0|^2 + \frac{2}{\varepsilon^{1/4}} \int_{I_\varepsilon} \hat{w}_0'\hat{\varphi}' \,. \quad (4.36)$$

As $\xi(s)$ is of class $\mathcal{C}^4([0,L])$, we may use the following Taylor expansion at fourth order in $\varepsilon^{1/4}$ (recall that $\nu_0^2 = \frac{1}{2}\rho_0\cdot\xi''(s_0)$ and $\bar{\xi}_\varepsilon(t) = \rho_0\cdot\xi(s_0 + \varepsilon^{1/4}t)$)

$$(\rho_0\cdot\bar{\xi}_\varepsilon)(t) - \mu_0 - \nu_0\sqrt{\varepsilon} = \varepsilon^{1/2}(\nu_0^2 t^2 - \nu_0) + \varepsilon^{3/4}\frac{t^3}{6}\rho_0\cdot\xi^{(3)}(s_0) + \varepsilon\frac{t^4}{24}\rho_0\cdot\xi^{(4)}(s_0) + o(\varepsilon) \,.$$

Then, taking into account the exponential decay of functions $\hat{w}_0, \hat{\varphi}$, the integrals over $I_\varepsilon$ in (4.36) can be substituted with the same integrals over all $\mathbb{R}$ and we obtain the following expansion:

$$U_\varepsilon(\hat{\varphi}) = \frac{1}{\sqrt{\varepsilon}} \int_\mathbb{R} (\nu_0^2 t^2 - \nu_0 + \frac{\nu_0}{2})|\hat{w}_0|^2 \, dt$$

$$+ \frac{1}{\varepsilon^{1/4}} \int_\mathbb{R} 2(\nu_0^2 t^2 - \nu_0)\,\hat{w}_0\,\hat{\varphi} + \frac{t^3}{6}(\rho_0\cdot\xi^{(3)}(s_0))\,|\hat{w}_0|^2 \, dt$$

$$+ \int_\mathbb{R} \left[ (\nu_0^2 t^2 - \nu_0)|\hat{\varphi}|^2 + \frac{2t^3}{6}(\rho_0\cdot\xi^{(3)}(s_0))\,\hat{w}_0\hat{\varphi} + \frac{t^4}{24}(\rho_0\cdot\xi^{(4)}(s_0)\,|\hat{w}_0|^2 \right] dt$$

$$+ \; o(\varepsilon^{1/4})$$

By (4.8) the first term in $\frac{1}{\sqrt{\varepsilon}}$ vanishes. Observing that $t^3|\hat{w}_0|^2$ and $\hat{w}_0\,\hat{\varphi}$ are odd functions, we see that the second term (in $\frac{1}{\varepsilon^{1/4}}$) vanishes as well and we conclude that

$$\lim_{\varepsilon\to 0} U_\varepsilon(\hat{\varphi}) = \int_\mathbb{R} \left[ (\nu_0^2 t^2 - \nu_0)|\hat{\varphi}|^2 + \frac{t^4}{24}(\rho_0\cdot\xi^{(4)}(s_0)\,|\hat{w}_0|^2 \right] dt \,.$$

The claim (4.31) follows from (4.35). □

# 5. APPENDIX

## 5.1. Some elementary results.

**Lemma 5.1.** *Let $A : H \to H$ be a positive linear self-adjoint operator with compact resolvent. Let $0 < \nu_0 < \nu_1 \leq \nu_2 \leq \cdots$ be the eigenvalues where the first one $\nu_0$ is assumed to be simple. Let $\{e_0, e_1, e_2, \cdots e_k \cdots\}$ be a basis of associated eigenvectors. Then:*
*i) For every $\lambda$, the following implication holds*

$$|\lambda - \nu_0| < \frac{\nu_1 - \nu_0}{2} \quad \Longrightarrow \quad \|Av - \lambda v\| \geq |\lambda - \nu_0| \|v\| \quad \forall v \in H .$$

*ii) For every $h \in H$ such that $(h|e_0) = 0$, there holds*

$$\min_{v \in H} \{(Av|v) - \nu_0 (v|v) + 2(h|v)\} = -\sum_{k=1}^{\infty} \frac{|(h|e_k)|^2}{\nu_k - \nu_0} .$$

*Proof.* The assertion i) is a consequence of the inequality $\|Av - \lambda v\| \geq \inf_{i \in \mathbb{N}} |\lambda - \nu_i|\} \|v\|$ (valid whenever $A$ is a self-adjoint operator). To show ii), we observe that, for every $v = \sum_k c_k e_k$, the energy $\mathcal{E}(v)$ to be minimized can be written as

$$\mathcal{E}(v) = \sum_{k=1}^{\infty} (\nu_k - \nu_0)|c_k|^2 + 2(h|e_k)c_k .$$

The minimum is achieved by taking $c_k = -\dfrac{(h|e_k)}{\nu_k - \nu_0}$ for $k \geq 1$ and $c_0$ arbitrary. □

## 5.2. The $\Gamma$-convergence method.

In this section we present a general result that enables us to guarantee the spectral convergence of our problem, throughout the $\Gamma$-convergence of the corresponding energy functional. The proof can be found in [2]. We begin by recalling the definition of $\Gamma$-convergence. Consider a quadratic functional $G : L^2(Q_L) \to (-\infty, +\infty]$. We say that the sequence $\{G_\varepsilon\}$ $\Gamma$-converges to $G$ in $H = L^2(Q_L)$ if the following two conditions hold:

(i) *(lower bound)* For any $v$ and $\{v_\varepsilon\}$ such that $v_\varepsilon \to v$ in $H$, $\liminf_{\varepsilon \to 0} G_\varepsilon(v_\varepsilon) \geq G(v)$;

(ii) *(upper bound)* For every $v$, there exists a sequence $\{\tilde v_\varepsilon\}$ such that $\tilde v_\varepsilon \to v$ in $H$ and $\limsup_{\varepsilon \to 0} G_\varepsilon(\tilde v_\varepsilon) \leq G(v)$.

It turns out that such a $\Gamma$-limit $G$ always exists, possibly after extracting a subsequence. Also, the $\Gamma$-convergence of $\{G_\varepsilon\}$ is unchanged if we subsitute $G_\varepsilon$ by its lower semicontinuous envelope (with respect to the strong topology in $H$) and the $\Gamma$-limit $G$ enjoys the lower semicontinuity property as well. For further features on $\Gamma$-convergence theory, we refer to the monograph by G. Dal Maso [5], where particular issues concerning the case of quadratic functionals and related linear operators are detailed (see Section 12 in this book). The relationship with the strong compact resolvent convergence of operators is summarized in the following

**Proposition 5.2.** *Let $A_\varepsilon : H_\varepsilon \to H_\varepsilon$ be a sequence of self-adjoint operators where $H_\varepsilon$ coincides algebraically with a fixed Hilbert space $H$ endowed with a scalar product $(\cdot|\cdot)_\varepsilon$ such that*

$$a_\varepsilon \|u\|^2 \leq (u|u)_\varepsilon \leq b_\varepsilon \|u\|^2,$$

*being $a_\varepsilon, b_\varepsilon$ suitable constants such that $a_\varepsilon, b_\varepsilon \to 1$ and $(\cdot, \cdot)$, $\| \cdot \|$ represent the usual scalar product and norm in $H$, respectively. Let $G_\varepsilon : H \to (-\infty, +\infty]$ be a lower semicontinuous quadratic form satisfying $G_\varepsilon(v) = (A_\varepsilon v|v)_\varepsilon$, if $v \in D(A_\varepsilon)$, and assume that the three following conditions hold:*

*i) $G_\varepsilon(v) \geq -c_0 \|v\|^2$ for a suitable constant $c_0 \geq 0$.*

*ii)* If $\sup_{\varepsilon}\{G_\varepsilon(v_\varepsilon) + \|v_\varepsilon\|\} < +\infty$, then $\{v_\varepsilon\}$ is strongly relatively compact in $H$.

*iii)* $G_\varepsilon$ does $\Gamma-$converge to $G$.

*Then, the limit functional $G$ determines a unique closed linear operator $A_0 : H \to H$ with compact resolvent, with domain $D(A_0)$ (a priori non dense in $H$) and such that $G(v) = (A_0 v, v)$ for all $v \in D(A_0)$. Furthermore, the spectral problems associated with $A_\varepsilon$ converge in the following sense: let $\mu_i^\varepsilon, v_i^\varepsilon$ and $\mu_i, v_i$ be such that*

$$v_i^\varepsilon \in H, \quad A_\varepsilon v_i^\varepsilon = \mu_i^\varepsilon v_i^\varepsilon, \quad \mu_0^\varepsilon \leq \mu_1^\varepsilon \leq \cdots \leq \mu_i^\varepsilon \cdots, \quad (v_\varepsilon^i | v_\varepsilon^j)_\varepsilon = \delta_{i,j},$$
$$v^i \in H, \quad A_0 v^i = \mu_i v_i, \quad \mu_0 \leq \mu_1 \leq \cdots \leq \mu_i \cdots, \quad (v_i | v_j) = \delta_{i,j}.$$

*Then, as $\varepsilon \to 0$, $\mu_i^\varepsilon \to \mu_i$ for every $i \in \mathbb{N}$. Moreover, up to a subsequence, $\{v_i^\varepsilon\}$ converges strongly to eigenvectors associated to $\mu_i$. Conversely, any eigenvector $v_i$ is the strong limit of a particular sequence of eigenvectors of $A_\varepsilon$ associated to $\mu_i^\varepsilon$.*

## Acknowledgements

M.L. Mascarenhas was partially supported by Fundação para a Ciência e a Tecnologia, through PEst-OE/MAT/UI0297/2011, PTDC/MAT109973/2009 and UTA-CMU/MAT/0005/2009. M.L. Mascarenhas thanks the hospitality of IMATH, Université du Sud-Toulon-Var, where part of this work was undertaken. L. Trabucho was supported by Fundação para a Ciência e a Tecnologia, through PEst-OE/MAT/UI0209/2011 and PTDC/MAT109973/2009.

Institut IMATH, Université du Sud-Toulon-Var, 83957 La Garde, Cedex, France
*E-mail address*: `bouchitte@univ-tln.fr`

Departamento de Matemática and Centro de Matemática e Aplicações, Faculdade de Ciências e Tecnologia, Universidade Nova de Lisboa, Quinta da Torre, 2829-516 Caparica, Portugal
*E-mail address*: `mascar@fct.unl.pt`

Departamento de Matemática, Faculdade de Ciências e Tecnologia, Universidade Nova de Lisboa, and CMAF, 2, Av. Professor Gama Pinto, 1649-003 Lisboa, Portugal
*E-mail address*: `trabucho@ptmat.fc.ul.pt`